\documentstyle[preprint,aps]{revtex}

\newcommand{\be}{\begin{equation}}
\newcommand{\ee}{\end{equation}}
\newcommand{\ba}{\begin{array}}
\newcommand{\ea}{\end{array}}

\tighten
\begin{document}

\draft
 
\title   {
Deformed Nuclear Halos
}
\author  {
T. Misu$^{1}$, 
 W. Nazarewicz$^{1-3}$,
and S. {\AA}berg$^{4,5}$
}

\address {
$^1$Department of Physics \& Astronomy,
 University of Tennessee,
  Knoxville, TN 37996, U.S.A.\\
$^2$Institute of Theoretical Physics, Warsaw University   
  Knoxville, TN 37996, U.S.A.\\
$^2$Institute of Theoretical Physics, Warsaw University         \\[-1mm]
   Ho\.za 69,  PL-00681, Warsaw, Poland                         \\
$^3$Physics Division, Oak Ridge National Laboratory             \\[-1mm]
   P.O. Box 2008, Oak Ridge,   TN 37831, U.S.A.                 \\
$^4$Joint Institute for Heavy Ion Research, 
   Oak Ridge National Laboratory                                \\[-1mm]
   P.O. Box 2008, Oak Ridge,   TN 37831, U.S.A.                 \\
$^5$Department of Mathematical Physics,
Lund Institute of Technology, S-22100 Lund, Sweden.
}
\maketitle
 
\begin{abstract}
Deformation properties of weakly bound nuclei
are discussed in the
deformed single-particle model.
 It is demonstrated that in the limit of a very
small binding energy the valence particles in specific orbitals,
characterized by a very small projection of single-particle angular
momentum onto the symmetry axis of a nucleus, can
give rise to the halo structure which is  completely decoupled from
the rest of the system. The quadrupole 
deformation of the resulting halo is completely
determined by the intrinsic structure of  a weakly bound orbital,
irrespective of the shape of the core.
\end{abstract}


\narrowtext

\section{Introduction}

The nature of exotic nuclei with extreme isospin values
 is one of the most exciting
challenges today, both experimentally and theoretically. 
Thanks to developments in experimental  technology \cite{[Gei95]}, 
we are in the process of  exploring the very limits of
nuclear existence, namely the regions
of the periodic chart in the neighborhood of the particle drip lines.
The systems of interest are characterized by extreme values of
isospin corresponding to large proton or neutron excess.

On the neutron-rich side, there
appears  a region of loosely bound 
 few-body systems,
neutron halos (see Refs.~\cite{[Mue93],[Rii94],[Han95],[Tan96]}
 for  reviews).
In these nuclei
the weak neutron binding implies large spatial dimensions and
the existence of the halo
(i.e., a dramatic excess of neutrons at large distances).
Theoretically, the weak binding and corresponding
closeness of the particle continuum, together with
the need for the explicit treatment of few-body dynamics,   
makes the  subject of halos both extremely interesting and difficult.

Neutron halos and heavy weakly bound  neutron-rich nuclei 
offer an opportunity to study the wealth of phenomena
associated with the closeness of the particle threshold: 
particle emission (ionization to the continuum) and  
characteristic behavior of cross sections \cite{[Wig48],[Fan61]},
 existence of soft 
collective modes  and low-lying transition strength
\cite{[Uch85],[Fay91],[Yok95],[Sag95],[Ham96]}, 
and dramatic changes in shell structure and
various nuclear properties in the sub-threshold regime 
\cite{[Ton78],[Hir91],[Dob94],[Dob96]}.

In this study  we address the notion of shape deformations in
halo nuclei. The importance of non-spherical
intrinsic shapes in halo nuclei  has been 
stressed in some papers,
especially in the context of a one-neutron halo $^{11}$Be and the nucleus
$^{8}$B, alleged to be a one-proton halo.
The ground state of  $^{11}$Be is a 1/2$^+$ state. The low neutron
separation energy, $S_n$=504 keV, allows for only one bound excited level
(1/2$^-$ at 320 keV). The halo character of $^{11}$Be 
has been confirmed by studies of reaction cross sections \cite{[Fuk91a]}
and the importance of  deformation can be inferred from the large
quadrupole moment of its  core
$^{10}$Be, $|Q|$=229\,mb  \cite{[Ram87]}.
The halo character of $^{8}$B has been suggested in Ref. \cite{[Min92]}
where the large measured quadrupole moment of its $I^\pi$=2$^+$ ground state,
$|Q|$=68\,mb, has been attributed to the weak binding of the fifth proton
in the $1p_{3/2}$ state. (The existence of proton halo in $^8$B is
still  heavily debated \cite{[Rii94],[Tan96]}. For 
instance,
Nakada and Otsuka \cite{[Nak94]}, in a shell-model calculation,
demonstrated that the large value of $|Q|$ in $^8$B could be
understood without introducing the proton halo.)

 The role of deformation in lowering the excitation energy of the  1/2$^+$ 
intruder  level  in $^{11}$Be has been recognized long ago. 
For instance,  Bouten {\it et al.}
 \cite{[Bou78]} pointed out that the position of the abnormal-parity
 intruder orbitals in odd $p$-shell nuclei  
can be  dramatically lowered by deformation,
 and they performed the projected Hartree-Fock  calculations 
 for the parity doublet in  $^{11}$Be.
In another paper, based on the cranked Nilsson model,
 Ragnarsson  {\it et al.} \cite{[Rag81]} demonstrated that 
 the parity doublet
could be naturally understood in terms of  the [220]1/2
($1d_{5/2}\otimes 2s_{1/2}$)  Nilsson orbital.  In particular,
they  calculated a very
large triaxial deformation for the positive-parity level, and 
a less-deformed prolate shape for the negative-parity state.
Muta and Otsuka \cite{[Mut95],[Ots95]} studied the structure of $^{11}$Be and
$^{8}$B with a deformed Woods-Saxon potential considering quadrupole 
deformation as a free parameter adjusted to the data. Tanihata
 {\it et al.} \cite{[Tan95]} concluded, based on a spherical
one-body potential, that the positions of experimental drip lines
are consistent with the spherical picture; they emphasized the
effect of increased binding of the low-$\ell$ shell model 
states near the threshold that can give rise to the level inversion.
In none of these above papers, however, have both effects, i.e.,
the loose binding
and self-consistency been simultaneously considered.

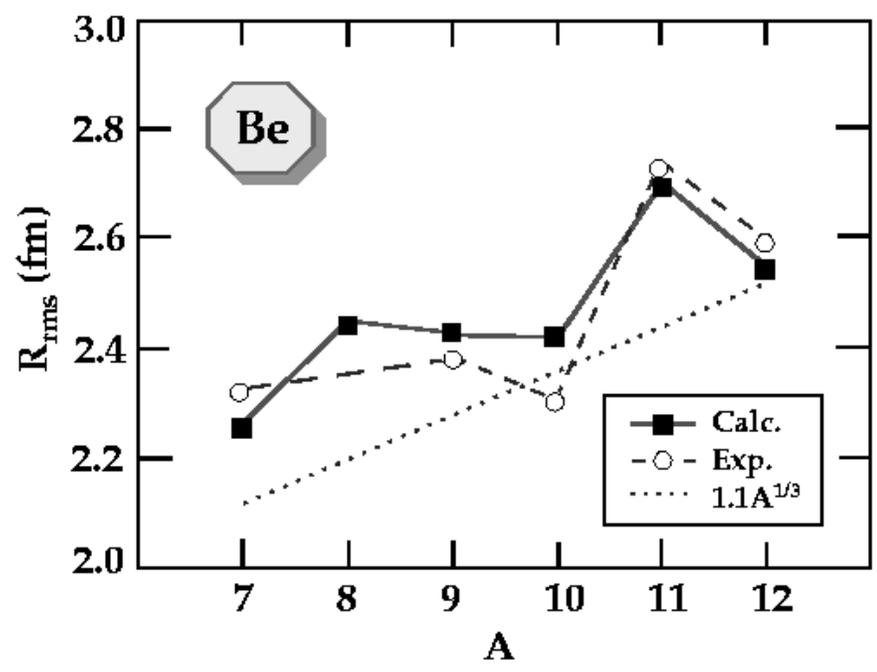
\begin{figure}[ht]
\caption{\label{Ba_radii}
Nuclear rms radii calculated in the modified
harmonic oscillator model (squares) for 
a series of
Be isotopes. They are compared to measured \protect\cite{[Tan88]}
 interaction radii (open 
circles). The calculated deviation from the smooth 
$1.1A^{1/3}$ behavior 
(dot-dash line) is due to deformation.
}
\end{figure}

Figure~\ref{Ba_radii} shows the
interaction radii
for a series of Be isotopes 
 deduced 
from measured interaction cross sections \cite{[Tan88]}. The 
relatively large radius for $^{11}$Be has been interpreted as a sign for 
halo structure of this nucleus. It is, however, quite interesting to note 
that calculated deformations (as obtained in Nilsson-Strutinsky 
calculations) of the Be isotopes are found to vary in such a way that 
the corresponding nuclear radii reproduce the data quite well.
In this case, no effects from the halo structure of $^{11}$Be have been 
considered since the calculations are based on the modified oscillator 
potential. Although these
 calculations are somewhat unrealistic, the result
displayed in Fig.~\ref{Ba_radii} 
clearly stresses the importance of simultaneously considering both
 deformation
and halo effects as, e.g., in $^{11}$Be.

Another, more microscopic, approach 
 is the work by Kanada-En'yo {\it et al.} 
\cite{[Kan95]} based on antisymmetrized
molecular dynamics with improved asymptotics.
 They obtained very large quadrupole
deformation for the  1/2$^+$ state
in $^{11}$Be, and a less deformed 
 1/2$^-$ state. A similar conclusion
has been drawn in the recent self-consistent Skyrme-Hartree-Fock 
calculations \cite{[Li96]}. 

Also, very recently,
the notion of deformation in  $^{11}$Be 
has been 
pursued by several authors 
\cite{[Esb95],[Vin95],[Ber95],[Nun96],[Rid96]}
within several variants of
the weak coupling scheme.
In these models, the odd neutron  moving in a Woods-Saxon potential 
is weakly coupled to the deformed core of 
$^{10}$Be,  and interacts with the core through the quadrupole field.
The
deformation
of $^{11}$Be is not treated self-consistently;
either the strength of the quadrupole
coupling is adjusted to the data to reproduce the quadrupole moment
of $^{10}$Be, or the deformation is adjusted
to reproduce the energies of the $I$=1/2 doublet.
The advantage of the weak coupling
approach is that the total
wave functions are eigenstates
of the total angular momentum and they have correct asymptotic
behavior. In this context, a nice molecular analogy,
in which the adiabatic coupling of the valence particle to the
deformed core is applied,  is
the weak binding of an electron to a rotationally excited
dipolar system \cite{[Gar71]}, or to a neutral symmetric molecule with
a non-zero quadrupole moment \cite{[Com96]}.

In this study,
we address the question of deformed two-body halos by considering
the single-particle motion in
 the axial spheroidal square well. The corresponding Schr\"odinger equation 
can be separated into three ordinary differential equations in the
spheroidal coordinate system. The properties of the deformed single-particle
states, especially in the subthreshold region, are analyzed by making
the multipole decomposition in the spherical partial waves 
with well-defined 
orbital angular momentum. Large spatial extensions of deformed halos
are discussed in terms of spherical radial form-factors.

The paper is organized as follows. Section
\ref{general} contains 
the discussion of the generic properties of deformed
halos. The method of evaluation
of the prolate spheroidal wave functions
is described in Sec.
\ref{model}, and the
 results of calculations are discussed in Sec.
\ref{results}. Finally, Sec.
\ref{conclusions} contains the main conclusions
of the paper.

\section{Deformed  neutron halos, general properties}\label{general}

This section contains some general arguments regarding the concept
of shape deformation in two-body halo systems. The 
material contained in this section 
is  the extension of the analysis of Ref.~\cite{[Rii92]}
carried out for the spherical case. (For the corresponding discussion
of three-body halo asymptotics, see Ref.\cite{[Fed93]}.) 

Let us assume that the weakly bound neutron moves in a deformed
average potential $U(\bbox{r})$ (usually well approximated
by a sum of the central potential and the spin-orbit potential).
The neutron wave function can be obtained by solving the deformed
single-particle Schr\"odinger equation
\be\label{deformedSE}
\left[\nabla^2 - \frac{2m}{\hbar^2}U(\bbox{r}) - \kappa_\nu^2 \right]
\psi_\nu(\bbox{r}) =0,
\ee
where $\kappa_\nu$=$\sqrt{-2m\epsilon_\nu/\hbar^2}$ and $\epsilon_\nu$ 
is the 
corresponding single-particle energy ($\epsilon_\nu$$<$0).

In the following considerations
we assume the axial reflection-symmetric 
central potential. (The generalization
to the triaxial and/or reflection-asymmetric
case is straightforward.) Since the asymptotic
properties
of radial matrix elements depend very weakly on intrinsic spin,
the spin-orbit term is neglected. 
Thanks to the axial symmetry, the single-particle states are labeled by
means of $\Lambda$ - the projection of the single-particle angular momentum
onto the symmetry axis ($z$-axis) and parity, $\pi$. Since the Hamiltonian 
considered is invariant with respect to the time-reversal
symmetry, we shall only consider 
non-negative values of $\Lambda$.

The deformed
wave function can be decomposed into spherical partial waves
with the well-defined orbital angular momentum $\ell$:
\be\label{multipole}
\psi^\Lambda_\nu(\bbox{r}) = \sum_{\ell} R_{\ell\Lambda\nu} (r) 
Y_{\ell\Lambda}(\hat{\bbox{r}}),
\ee
where, due to the assumed reflection symmetry,
\be\label{parity}
\pi = (-1)^{\ell}.
\ee
At large distances ($r$$>$$R$, where $R$ defines the 
arbitrary but fixed distance
at which the nuclear 
interacion becomes unimportant),  $U(\bbox{r}$) vanishes and
the
radial functions $R_{\ell\nu} (r)$ satisfy the free-particle
radial Schr\"odinger equation
\be\label{sphericalSE}
\left[\frac{d^2}{dr^2} + \frac{2}{r}\frac{d}{dr} -\kappa_\nu^2 - 
\frac{\ell(\ell+1)}{r^2}\right] R_{\ell\nu} (r)=0.
\ee
(Of course, $R_{\ell\Lambda\nu}(r)$=$R_{\ell\nu} (r)$ for $r\geq R$.) 
The solution of Eq.~(\ref{sphericalSE}) is $R_{\ell\nu}(r)=
B_{\ell} h^+_{\ell}(i\kappa_\nu r)$
where $h^+_{\ell}$ is the spherical Hankel function
and $B_{\ell}$ is a constant, given by 
\be\label{BB}
B_{\ell} = R_{\ell\nu}(R)/h^+_{\ell}(i\kappa_\nu R).
\ee

The spatial properties of the system can be characterized  by
radial moments,
$\langle \psi^\Lambda_\nu| r^n|\psi^\Lambda_\nu\rangle$, 
and  multipole moments
$\langle \psi^\Lambda_\nu| r^n Y_{n0}|\psi^\Lambda_\nu\rangle$.
Both quantities require the evaluation of radial matrix elements
\be\label{rme}
\langle\ell\Lambda\nu|r^n|\ell'\Lambda\nu\rangle \equiv 
\int_{0}^{\infty} r^{n+2}
R^*_{\ell\Lambda\nu}(r)R_{\ell'\Lambda\nu}(r) dr =
I_{n\ell\ell'\Lambda\nu}+ O_{n\ell\ell'\nu},
\ee
where $I$ represents the contribution from the inner ($r$$<$$R$)
region and $O$ is the outer region contribution
($r$$>$$R$).
Thanks to parity  conservation [Eq.~(\ref{parity})], 
$\ell'=\ell \bmod 2$. 
 The inner integral
is, by definition, finite. The outer integral can be written as
\begin{eqnarray}\label{asymptote}
O_{n\ell\ell'\nu} & = & \int_{R}^{\infty} r^{n+2} B^*_{\ell} B_{\ell'}
h^{+*}_{\ell}(i\kappa_\nu r) h^+_{\ell'}(i\kappa_\nu r) dr \\
 & = & B^*_{\ell} B_{\ell'} \kappa_\nu^{-(n+3)}
\int_{R\kappa_\nu}^{\infty} h^{+*}_{\ell}(ix) h^+_{\ell'}(ix)
x^{n+2} dx.
\end{eqnarray}
In the limit of a very weak binding ($\kappa_\nu\rightarrow 0$)
one can replace the value of $h^+_{\ell}(i\kappa_\nu R)$ 
in Eq.~(\ref{BB})
by the asymptotic expression valid for small arguments. This gives:
\be
B_{\ell} \approx \frac{i^{\ell+1}}{1\times 3\times...(2\ell -1)}
R_{\ell\nu}(R) (R\kappa_\nu)^{\ell +1}.
\ee
Now, following the
arguments of Ref.~\cite{[Rii92]}, one can demonstrate that 
for small values of $\kappa_\nu$,
$O_{n\ell\ell'\nu}$ behaves asymptotically as
\be\label{finale}
O_{n\ell\ell'\nu} \propto
\kappa_\nu^{\ell + \ell' - n - 1}\left\{
\frac{{x_0}^{n-\ell-\ell'+1} -
{(R\kappa_\nu)}^{n-\ell-\ell'+1}}{n-\ell-\ell'+1}
+ \text{const.} \right\},
\ee
where $x_0\gg R\kappa_\nu$ is a small constant. Consequently,
the asymptotic behavior of $O_{n\ell\ell'\nu}$ 
in  the limit of small $\epsilon$ strongly depends on quantum
numbers $n$, $\ell$, and $\ell'$. Namely, for
\begin{eqnarray}\label{conditions}
n> \ell + \ell' -1: & O_{n\ell\ell'\nu} &  ~\text{diverges as} 
~(-\epsilon_\nu)^{(\ell + \ell' -n -1)/2},\label{c1} \\
n=  \ell + \ell' -1: &  O_{n\ell\ell'\nu} & ~\text{diverges as}
~-\frac{1}{2}\ln(-\epsilon_\nu),\label{c2}\\
n< \ell + \ell' -1:  & O_{n\ell\ell'\nu} & 
~\text{remains finite} \label{c3}.
\end{eqnarray}

\subsection{The normalization integral}\label{nint} 

The norm of the deformed state [Eq~(\ref{multipole})],
${\cal N}_{\Lambda\nu}$,
 can be expressed through
the zeroth radial moment
\be\label{norm}
({\cal N}_{\Lambda\nu})^2=\langle\psi^\Lambda_\nu|\psi^\Lambda_\nu\rangle =
\sum_{\ell}\langle\ell\Lambda\nu|r^0|\ell\Lambda\nu\rangle=\sum_{\ell}
\left(I_{0\ell\ell\Lambda\nu}+ O_{0\ell\ell\nu}\right).
\ee
Consequently, according to Eq.~(\ref{conditions}),
the norm is divergent only if the deformed state contains an
admixture of the $s$-wave. This is possible  only for 
orbitals with
$\pi$=+ and  $\Lambda$=0. In this case, the norm
behaves asymptotically as
$\sqrt{O_{000\nu}}$$\propto$$(-\epsilon_\nu)^{-1/4}$, and the probability
to find the neutron in the outer region,
\be\label{outer}
P_{\rm outer}=\frac{O}{I+O},
\ee
approaches one for zero binding.

\subsection{The rms radius}

The root-mean-square radius of  a deformed orbital is given by
\be\label{radius}
\langle \Lambda\nu| r^2| \Lambda\nu\rangle \equiv
\frac{\langle\psi^\Lambda_\nu|r^2|\psi^\Lambda_\nu\rangle}
{\langle\psi^\Lambda_\nu|\psi^\Lambda_\nu\rangle}
 = ({\cal N}_{\Lambda\nu})^{-2}
\sum_{\ell}\langle\ell\Lambda\nu|r^2|\ell\Lambda\nu\rangle.
\ee
As discussed in Ref.~\cite{[Rii92]}, with decreasing binding energy
the integral $O_{2\ell\ell\nu}$ diverges  as 
$(-\epsilon_\nu)^{-3/2}$ for $\ell$=0 and 
as $(-\epsilon_\nu)^{-1/2}$ for $\ell$=1 [see Eq.~(\ref{conditions})].
Therefore, in the deformed system, the rms radius  diverges
only if the Nilsson  orbital in question
contains a
component of an
$s$ or a  $p$ state. This leaves only three classes
of states for which the spatial extension can be
arbitrary large:
\begin{eqnarray}\label{conditions_r}
\pi=+, \Lambda=0:~~~  & \langle r^2 \rangle & ~\text{diverges as} 
~(-\epsilon_\nu)^{-1},\nonumber \\
\pi=-, \Lambda=0~\text{or}~1:  & \langle r^2 \rangle & 
~\text{diverges as} 
~(-\epsilon_\nu)^{-1/2}.
\end{eqnarray}
In the following, these states are referred to as ``halo states"
or ``halos". Of course, this does not mean that other
Nilsson orbitals cannot form very extended structures when their
binding becomes very small. However, it is only for the 
states (\ref{conditions_r}) that the rms radius becomes infinite
asymptotically.

\subsection{The quadrupole moment}

The average quadrupole moment of a deformed orbital is given by
\be\label{quadrupole}
\langle \Lambda\nu| r^2 Y_{20}| \Lambda\nu\rangle =
\frac{\langle\psi^\Lambda_\nu|r^2Y_{20}|\psi^\Lambda_\nu\rangle}
{\langle\psi^\Lambda_\nu|\psi^\Lambda_\nu\rangle}
 = ({\cal N}_{\Lambda\nu})^{-2}
\sum_{\ell\ell'}\langle\ell\Lambda\nu|r^2|\ell'\Lambda\nu\rangle 
\langle\ell\Lambda|Y_{20}|\ell'\Lambda\rangle,
\ee
where the angular matrix element is
\be\label{CG}
\langle\ell\Lambda|Y_{20}|\ell'\Lambda\rangle
= \sqrt{\frac{5}{4\pi}}\sqrt{\frac{2\ell' +1}{2\ell +1}}
\langle \ell'\Lambda 2 0|\ell\Lambda\rangle
\langle \ell'0  2 0|\ell 0\rangle.
\ee
According to Eq.~(\ref{c1}), in the $\epsilon_\nu\rightarrow 0$ limit
the 
quadrupole matrix element diverges
if  $\ell+\ell'< 3$.
Since the quadrupole moment of an $s$ state vanishes, the only
diverging matrix element 
among the $\pi$=+ states
comes from an $s$$\leftrightarrow$$d$
coupling.
At small binding energies,
the corresponding integral $O_{202\nu}$ behaves as 
$(-\epsilon_\nu)^{-1/2}$.
For negative-parity orbitals, the only diverging matrix element
is the diagonal one, $O_{211\nu}$, 
which also behaves as $(-\epsilon_\nu)^{-1/2}$.

However, because of different asymptotic properties
of the normalization integrals,
the low-$\epsilon$ behavior 
of the single-particle quadrupole moment of a 
weakly bound  orbital
does depend on parity. 
With $\epsilon_\nu$$\rightarrow$0, 
 the quadrupole
moment (\ref{quadrupole})
of the $\pi$=+ halo
 approaches the finite limit.
On the other hand, 
for the $\pi$=-- halos the norm of the state remains finite and 
the quadrupole moment
behaves as $(-\epsilon_\nu)^{-1/2}$.

It is instructive to  consider
the quadrupole deformation $\beta_2$ extracted from 
the ratio 
\be\label{beta}
\beta_2 \equiv \frac{4\pi}{5} \frac{\langle r^2 Y_{20}\rangle}
{\langle r^2 \rangle}.
\ee
By splitting ${\langle r^2 Y_{20}\rangle}$ and ${\langle r^2 \rangle}$
into contributions from the core (c) and from the valence (v)
nucleons,
one obtains
\be\label{b11}
\beta_2 = \frac{4\pi}{5} \frac{\langle r^2 Y_{20}\rangle_{\rm c}
+ \langle r^2 Y_{20}\rangle_{\rm v}}
{\langle r^2 \rangle_{\rm c}+ \langle r^2 \rangle_{\rm v}}.
\ee

For positive-parity halos ($\pi$=+, $\Lambda$=0),
the numerator in Eq.~(\ref{b11}) is finite while the denominator 
diverges as $(-\epsilon_\nu)^{-1}$. Hence
$\beta_2$ is asymptotically linear in $\epsilon_\nu$, i.e.,
it {\em vanishes} in the limit of zero binding:
\be\label{bpos}
\beta_2 (\pi=+, \Lambda=0)\stackrel{\epsilon_\nu\rightarrow 0}
{\longrightarrow}
0.
\ee
 On the other hand,
for negative-parity halos ($\pi$=--, $\Lambda$=0 or 1), 
the  ratio (\ref{b11}) is solely determined by the
the $p$-wave components in the valence state:
\be\label{bneg}
\beta_2 (\pi=-, \Lambda) \stackrel{\epsilon_\nu\rightarrow 0}
{\longrightarrow}
\frac{4\pi}{5}
\langle 1\Lambda|Y_{20}|1\Lambda\rangle =
\left\{
\begin{array}{ll}
    +0.63 & \text{if}~\Lambda=0 \\
    -0.31 & \text{if}~\Lambda=1.
\end{array}
\right.
\ee
That is, the deformation of the halo is solely determined by
the spatial structure of the valence state wave function,
independently of the shape of the core. The deformed
core merely establishes  the quantization axis of the system, 
important for determining the projection $\Lambda$.

\subsection{Higher moments and  multipole deformations}
The above discussion is easily generalized to the case of
higher multipoles. For instance, for $n$=4,  the hexadecapole moment
$\langle r^4 Y_{40}\rangle$
behaves asymptotically in the same manner as the quadrupole moment, i.e.,
it approaches the finite limit for the $\pi$=+ halos and diverges as
$(-\epsilon_\nu)^{-1/2}$ for the $\pi$=-- halos.
However, the corresponding deformation $\beta_4$, proportional
to $\langle r^4 Y_{40}\rangle/\langle r^4 \rangle$ approaches zero,
regardless of  parity of the halo orbital.

\section{The Model}\label{model}

In our study,
the deformed potential $U(\bbox{r})$ 
has been  approximated by a prolate
spheroidal finite square well potential. 
Spheroidal {\em infinite} square well
was early used by
Moszkowski \cite{[Mos55]} to discuss the properties of single-particle
deformed orbitals. 
Merchant and Rae \cite{[Mer94]}
investigated the continuum spectrum ($\epsilon$$>$0) of the
spheroidal finite  square well potential
  to calculate the particle
decay widths of deformed nuclei. 
 Since  the main focus of our work  is  the behavior
of bound single-particle orbitals very close to the 
$\epsilon$=0 threshold, particular
attention was paid to a  precise numerical solution of
the Schr\"odinger equation in the limit of very small binding 
energies and/or large deformations.

\subsection{Prolate
Spheroidal Coordinates and Parametrization of Nuclear Shape}

Assuming the $z$-axis to be a symmetry axis of the nucleus,
 the coordinate transformation between the prolate spheroidal
coordinates ($\xi$, $\eta$, $\phi$) and the cartesian coordinates 
($x$,$y$,$z$) reads  \cite{[Mor53],[Lit56],[Abr65]}:
\begin{eqnarray}\label{coord}
    x & = & a\sqrt{(\xi^2-1)(1-\eta^2)}\cos\phi ,\label{coordx} \\
    y & = & a\sqrt{(\xi^2-1)(1-\eta^2)}\sin\phi  ,\label{coordy}\\
    z & = & a\xi\eta ,\label{coordz} 
\end{eqnarray}
where $a>0$,
 $1\leq\xi\leq\infty$, $-1\leq\eta\leq 1$, and $0\leq\phi\leq 2\pi$.

Surfaces of constant $\xi$=$\xi_0$ represent confocal ellipses,
\be\label{ellipse}
\frac{x^2+y^2}{a^2(\xi_0^2-1)} + \frac{z^2}{a^2\xi_0^2} = 1,
\ee
with foci at
(0,0,$\pm$$a$),
minor axis $R_\perp$=$a\sqrt{\xi_0^2-1}$, and major axis 
$R_\parallel$=$a\xi_0$.
(Since the purpose of this study
is to investigate the generic features of weakly bound states
in a deformed potential, the analysis is limited to prolate 
shapes.
However, 
the calculations can easily be extended to the oblate side
 through a simple coordinate transformation.) 
 
It is seen from Eq.~(\ref{ellipse}) that
the parameter $\xi_0$ defines the 
shape deformation of a system. Indeed,
for $\xi_0\gg1$, the surface (\ref{ellipse}) becomes that of
a sphere with the radius $a\xi_0$, while the limit $\xi_0\rightarrow 1$
corresponds to  a segment. Following
Ref.~\cite{[Mos55]}, we
introduce the deformation parameter $\delta$:
\be\label{delta}
\delta=\left(\frac{R_\parallel}{R_\perp}
\right)^{2/3}-1= \left(\frac{\xi^2_0}{\xi^2_0-1}\right)^{1/3}-1.
\ee
The volume-conservation condition
[the volume inside  the surface (\ref{ellipse}) should not
 depend on $\delta$]
yields
\be\label{RRR}
a = \frac{R_0}{(\xi_0^3-\xi_0)^{1/3}},
\ee
where $R_0$ is the corresponding spherical radius. 

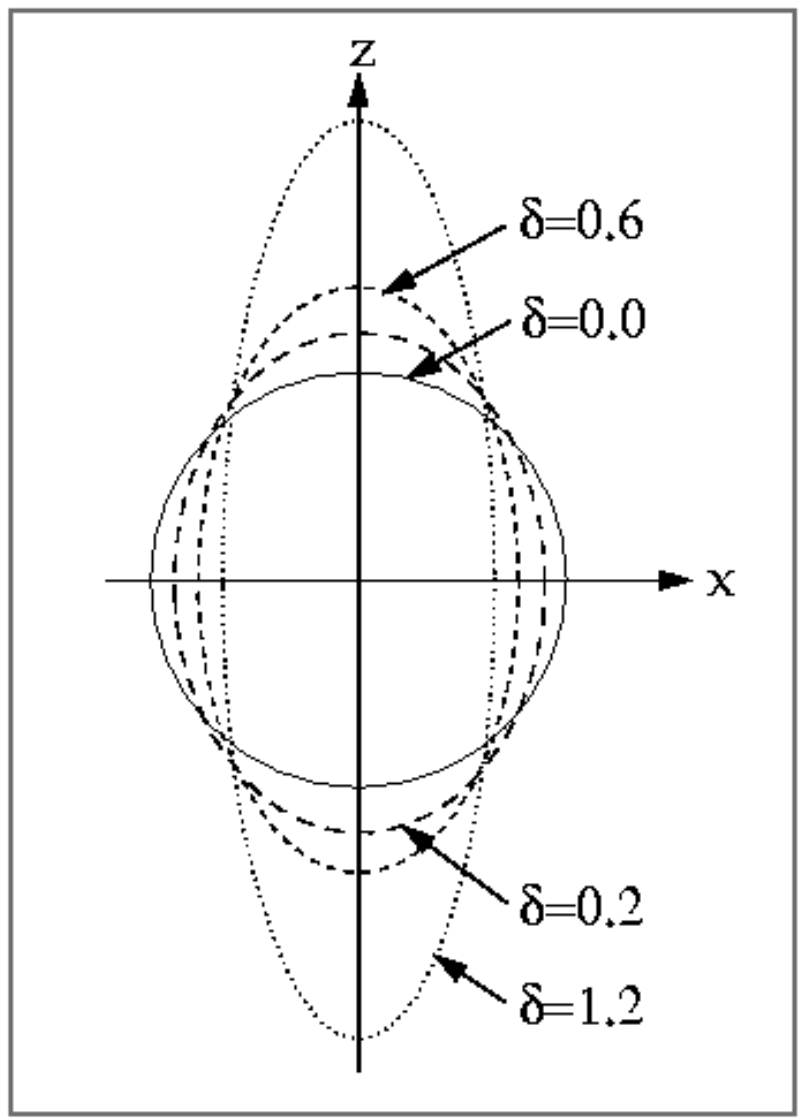
\begin{figure}[ht]
\caption{\label{shapes}
Shapes corresponding to quadrupole deformations
$\delta$=0, 0.2, 0.6, and 1.2, assuming the volume conservation
condition, Eq.~(\protect\ref{RRR}). 
}
\end{figure}

To find the relation between $\delta$ and other
quadrupole  deformation parameters, one can compare the
 macroscopic quadrupole moment  of the surface
  (\ref{ellipse})
\be\label{Qe}
Q_2 = \sqrt{\frac{16\pi}{5}}\langle r^2Y_{20}\rangle
=  \frac{2}{5}R_0^2 \delta\frac{\delta^2+3\delta+3}{\delta+1}
\ee
with those obtained using other shape parametrizations \cite{[Naz96a]}.
For example, the relation between $\delta$ and the oscillator
deformation $\delta_{\rm osc}$ is
\be\label{deltaosc}
\delta=\left(\frac{1+{1\over 3}\delta_{\rm osc}}
{1-{2\over 3}\delta_{\rm osc}}\right)^{2/3}-1.
\ee
For small values of $\delta_{\rm osc}$, Eq.~(\ref{deltaosc})
gives $\delta = \frac{2}{3}\delta_{\rm osc}$. However, at a 
superdeformed shape ($\frac{R_\parallel}{R_\perp}$=2), both
 deformations are 
very close: $\delta_{\rm osc}$=0.6 
while   $\delta$=2$^{2/3}$--1=0.587.
Figure~\ref{shapes} shows the family of shapes representing different
deformations $\delta$.

\subsection{Bound States in the Prolate Spheroidal Well}

The  deformed spheroidal square well potential is given by
\be\label{pot}
U(\xi) =
\left\{
\begin{array}{ll}
    U_{0} & \text{for}~\xi\leq\xi_{0} \\
    0         & \text{for}~\xi>\xi_{0},
\end{array}
\right.
\ee
where $U_0$ is the depth of the potential well 
($U_0$$<$0) and $\xi_0$
depends on $\delta$ through
Eq.~(\ref{delta}).

Expressed in  prolate spheroidal
coordinates, 
the time-independent Schr\"{o}dinger equation (\ref{deformedSE})
can be written as
\begin{eqnarray}\label{schrod}
\left[\frac{\partial}{\partial\xi}\left\{(\xi^2-1)
\frac{\partial}{\partial\xi}\right\}\right.
& + &
\left.
\frac{\partial}{\partial\eta}\left\{(1-\eta^2)
\frac{\partial}{\partial\eta}\right\}
+\frac{\xi^2-\eta^2}{(\xi^2-1)(1-\eta^2)}
\frac{\partial^2}{\partial\phi^2}\right]
\psi(\xi,\eta,\phi) \nonumber\\
 & + & \frac{2ma^2(\eta^2-\xi^2)}{\hbar^2}\left[U(\xi)-\epsilon\right]
\psi(\xi,\eta,\phi) = 0.
\end{eqnarray}
Following Ref.~\cite{[Mor53]},
this equation can be separated into three ordinary
differential equations 
by  assuming $\psi(\xi,\eta,\phi)=R(\xi)S(\eta)\Phi(\phi)$.
The functions $R$, $S$, and $\Phi$ are solutions of the
ordinary differential equations
\begin{eqnarray}\label{schrodsep}
\frac{d}{d\xi}
\left[(\xi^2-1)\frac{dR_{\Lambda l}(c,\xi)}{d\xi}\right] 
&-& 
\left[\lambda_{\Lambda l}-c^2\xi^2+\frac{\Lambda^2}{\xi^2-1}\right]
R_{\Lambda l}(c,\xi)
 =  0 ,\label{schrod1} \\ 
\frac{d}{d\eta}
\left[(1-\eta^2)\frac{dS_{\Lambda l}(c,\eta)}{d\eta}\right] 
& + & \left[\lambda_{\Lambda l}-c^2\eta^2-\frac{\Lambda^2}{1-\eta^2}\right]
S_{\Lambda l}(c,\eta) =  0 ,\label{schrod2}\\ 
\frac{d^2\Phi_\Lambda(\phi)}{d\phi^2} & = & -\Lambda^2
\Phi_\Lambda(\phi),\label{schrod3}
\end{eqnarray}
where   $\lambda_{\Lambda l}$ is the separation constant and
\be\label{c22}
c =
\left\{
\begin{array}{ll}
 c_{\rm int} = a\sqrt{2m(\epsilon-U_0)}/\hbar & \text{for}~\xi\leq\xi_0 \\
 ic_{\rm ext} = ia\sqrt{-2m \epsilon}/\hbar & \text{for}~\xi>\xi_0.
\end{array}
\right.
\ee
In the following, 
 $R_{\Lambda l}(c,\xi)$ and $S_{\Lambda l}(c,\eta)$ 
are referred to as
 the radial and angular
spheroidal functions, respectively.
For positive values of $\epsilon$, scattering solutions for the spheroidal
square well were solved in Ref.~\cite{[Mer94]}. Here, we concentrate on
bound states with $\epsilon<0$.

The angular solution $S_{\Lambda l}(c,\eta)$ can be expressed in terms
of a series of  the associated 
Legendre
functions of the first  kind
\be\label{angwave1}
    S_{\Lambda l}(c,\eta)   =  {\sum_{k}^{\infty}}'
d^{\Lambda l}_k(c)P^\Lambda_{\Lambda+k}(\eta),
\ee
where the prime over the summation sign  indicates
that $k$=0, 2, $\ldots$ if ($l-\Lambda$) is even, and
$k$=1, 3, $\ldots$ if ($l-\Lambda$) is odd \cite{[Mor53],[Abr65]}
(parity conservation). 

The radial functions   $R_{\Lambda l}(c,\xi)$ 
 are expanded in terms of spherical Bessel functions
of the first kind,
$f^{(1)}_k(c\xi)\equiv j_k(c\xi)$, 
and spherical Bessel functions of the third kind 
$f^{(3)}_k(c\xi)\equiv h_k(c\xi)=j_k(c\xi)+in_k(c\xi)$. 
The internal radial function $R^{(1)}_{\Lambda l}(c,\xi)$
 ($\xi\le\xi_0$)
and the external  radial function
$R^{(3)}_{\Lambda l}(c,\xi)$
 ($\xi >\xi_0$) can be written as
\be\label{radwave}
R^{(p)}_{\Lambda l}(c,\xi)
=\left\{{\sum_{k}^{\infty}}'\frac{(2\Lambda+k)!}{k!}
d^{\Lambda l}_k(c)\right\}^{-1}
\left(\frac{\xi^2-1}{\xi^2}\right)^{\Lambda/2}
{\sum_{k}^{\infty}}'
i^{k+\Lambda- l}\frac{(2m+k)!}{k!}d^{\Lambda l}_k(c)
f^{(p)}_{\Lambda+k}(c\xi)
\ee
where $p$=1 or 3.

Finally, the 
deformed single-particle wave function is given by
\be\label{waveint}
\psi_{\Lambda n_{\rm exc}}(\xi,\eta,\phi) =
\left\{
\begin{array}{ll}
\sum_{l}^{\infty}A_{n_{\rm exc}\Lambda l}
R^{(1)}_{\Lambda l}(c_{\rm int},\xi)
S_{\Lambda l}(c_{\rm int},\eta)\Phi_\Lambda(\phi) &
\mbox{for $\xi\leq\xi_0$} \\
\sum_{l}^{\infty}B_{n_{\rm exc}\Lambda l}
R^{(3)}_{\Lambda l}(ic_{\rm ext},\xi)
S_{\Lambda l}(ic_{\rm ext},\eta)\Phi_\Lambda(\phi) &
\mbox{for $\xi>\xi_0$} 
\end{array}
\right.
\ee
where $c_{\rm int}$ and $c_{\rm ext}$ are defined
 in  Eq.~(\ref{c22}),
and $n_{\rm exc}$ is the excitation  quantum number
labeling   orbitals
having  the same quantum numbers $\Lambda$ and $\pi$=$(-1)^l$.

By matching the internal and external wave functions at  $\xi$=$\xi_0$, 
one finds the eigenenergies $\epsilon$ and the amplitudes
$A_{n_{\rm exc}\Lambda l}$ and $B_{n_{\rm exc}\Lambda l}$.
 The details of the calculation are outlined in 
 Appendix~\ref{appA}. The procedure used to calculate
the separation constant
$\lambda_{\Lambda l}$ and
coefficients $d^{\Lambda l}_k(c)$ is discussed in 
Appendix~\ref{appB}.

\section{Results}\label{results}

This section  illuminates the general properties
of deformed Nilsson orbitals discussed in Sec.~\ref{general}
using the spheroidal square well potential. 

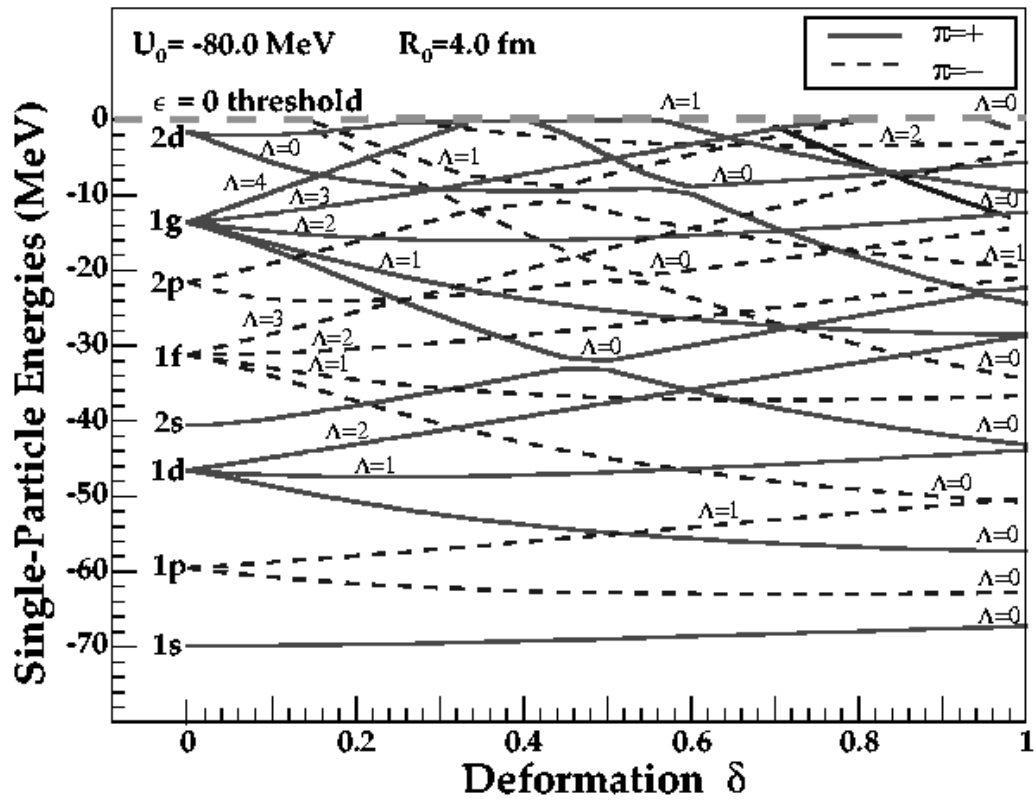
\begin{figure}[ht]
\caption {\label{sp}
Single-particle energies of the finite
spheroidal well with $U_0$=--80\,MeV and $R_0$=4\,fm 
as a function of deformation $\delta$. At a spherical shape the orbitals are
characterized
by means of  quantum numbers ($n\ell$). 
The deformed orbitals are labeled by parity 
($\pi$=+, solid line; 
$\pi$=--, dashed line) and by 
orbital angular momentum
projection onto the symmetry axis ($z$-axis), $\Lambda$.
}
\end{figure}

The single-particle energies of the finite
spheroidal well with $U_0$=--80\,MeV and $R_0$=4\,fm 
are shown in Fig.~\ref{sp}
as  functions
 of deformation $\delta$. At a spherical shape the orbitals are
characterized
by means of spherical quantum numbers ($n\ell$). 
The deformed orbitals are labeled by parity $\pi$, angular momentum
projection $\Lambda$, and the excitation quantum number $n_{\rm exc}$
which specifies the energetic order of a single-particle orbital in a given
($\pi\Lambda$)-block, counting from the bottom of the well
(e.g., $n_{\rm exc}$=1 corresponds to the lowest state, 
$n_{\rm exc}$=2 is the second state, and so on).
In the following, the  deformed orbitals are labeled as 
[$n_{\rm exc}\Lambda\pi$]. 
 For example, the $\Lambda$=1 orbital
originating from the spherical shell $1d$ is referred to as [11+]
(see Fig.~\ref{sp}).

\subsection{Radial Properties of Deformed Orbitals}

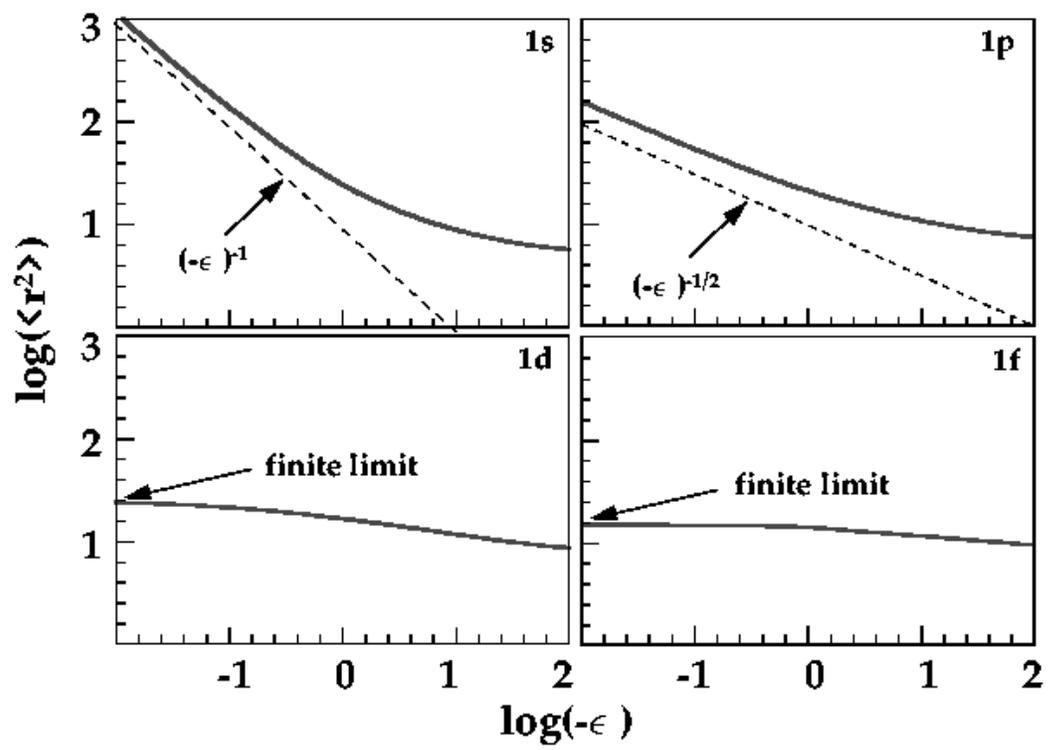
\begin{figure}[ht]
\caption {\label{rads}
Dependence of the single-particle rms radius on binding energy
$\epsilon$
at $\delta$=0 (spherical shape) and  $R_0$=4\,fm. The potential
depth is adjusted in each case to obtain the desired value of $\epsilon$.
The values of $\langle r^2 \rangle$ for $1s$ and $1p$ orbitals
diverge at small binding energies according to the power
law, Eq.~(\protect\ref{conditions_r}).
}
\end{figure}

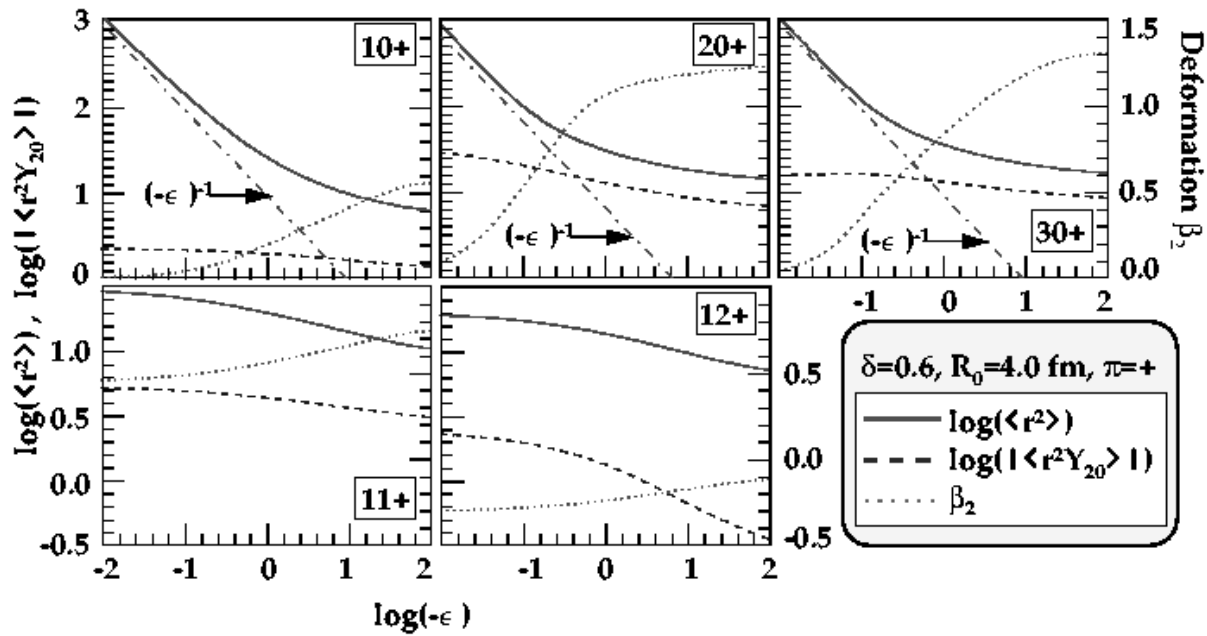
\begin{figure}[ht]
\caption {\label{raddp}
Binding energy dependence of $\langle r^2\rangle$ (solid line), 
$\langle r^2Y_{20}\rangle$  (dashed line), and $\beta_2$ (dotted line)
for several $\pi$=+ orbitals
at $\delta$=0.6 (superdeformed shape).
The asymptotic limits discussed in Sec.~\protect\ref{general} are indicated.
}
\end{figure}

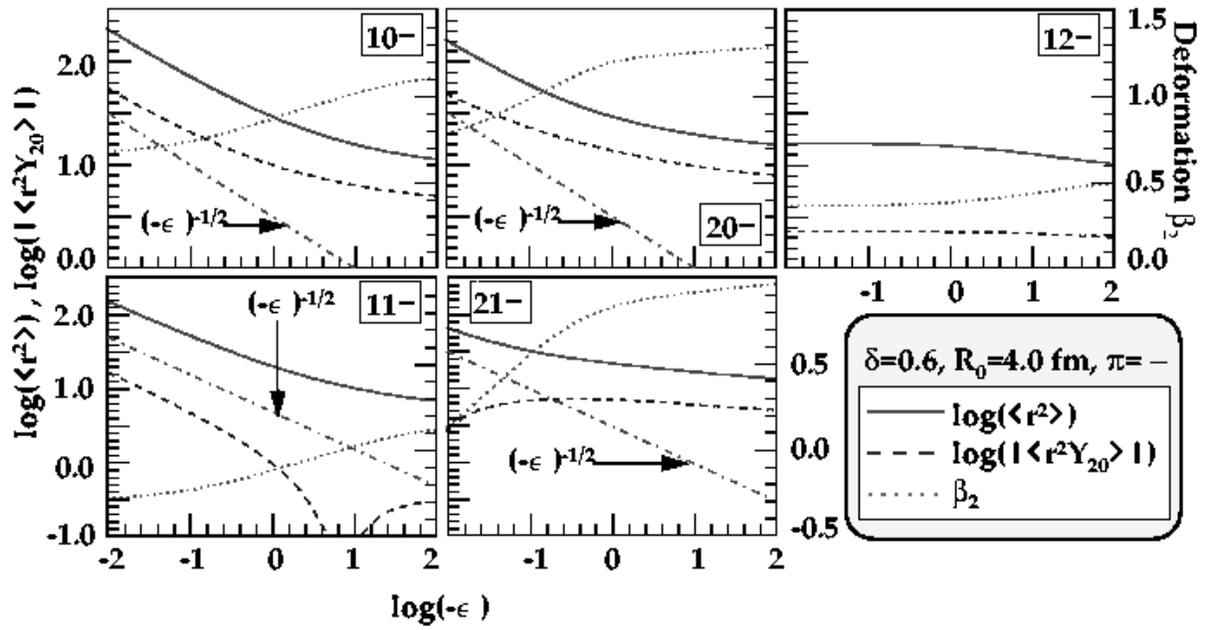
\begin{figure}[ht]
\caption {\label{raddn}
Same as in Fig.~\protect\ref{raddp} except
for the $\pi$=-- states.
}
\end{figure}

The dependence of the single-particle rms radius on binding energy
is illustrated in Fig.~\ref{rads} (spherical shape), and 
Figs. \ref{raddp} and
\ref{raddn}
 (superdeformed shape).
(In calculations, the  binding energy was varied by changing
  the well depth $U_0$.)

 The spherical case has been discussed
in detail in Ref.~\cite{[Rii92]}; here it is shown for the
mere reference only. In all cases the
asymptotic conditions [Eq.~(\ref{conditions_r})] are met rather quickly.
Indeed, in the considered range of binding energy the values of 
$\langle r^2 \rangle$ for the $1s$ state 
shown in Fig.~\ref{rads} and 
the [10+], [20+], and [30+] orbitals  of Fig.~\ref{raddp} approach
an asymptotic limit [$(-\epsilon)^{-1}$ dependence], and similar
 holds for
the $1p$ state and [10--], [20--], [11--], and [21--] orbitals
(see Fig.~\ref{raddn}) which
behave as $(-\epsilon)^{-1/2}$ at low binding energy. The remaining
states do not exhibit any halo effect, as expected.

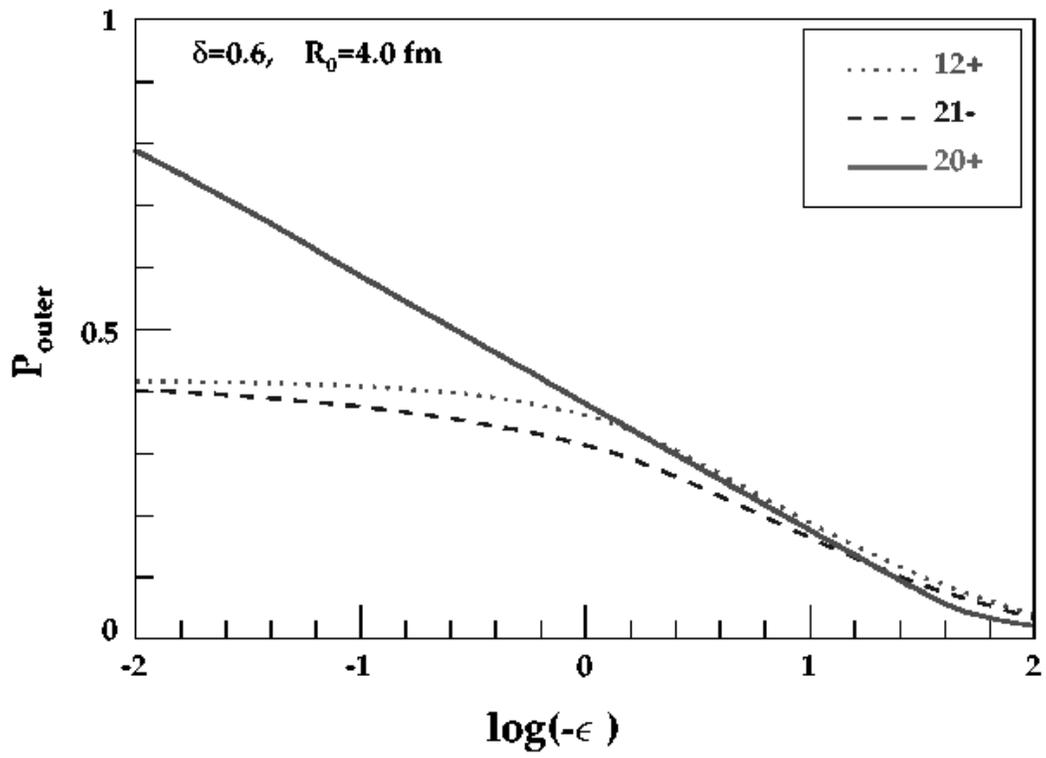
\begin{figure}[ht]
\caption {
Probability 
$P_{\rm outer}$, Eq.~(\protect\ref{outer}), to
find the neutron in the classically forbidden region, $\xi > \xi_0$,
as a function of $\epsilon$
for three superdeformed states [12+], [21--], and [20+]. 
}\label{out}
\end{figure}

Figure \ref{out} displays 
the probability 
$P_{\rm outer}$ [Eq.~(\ref{outer})]  to 
find the neutron in the classically forbidden region, $\xi > \xi_0$,
as a function of $\epsilon$ 
for three superdeformed states with different values of $\Lambda$.
At low values of binding energy, 
the 
$\ell$=0 component  completely dominates the structure of the 
[20+]  state
and $P_{\rm outer}\rightarrow 1$.

The
radial form factors $R_{\ell\Lambda n_{\rm exc}}(r)$ appearing in
 the multipole
decomposition [Eq.~(\ref{multipole})] carry information about the spatial
extension of the wave function. They can be obtained by the angular
integration:
\be\label{ffactors}
R_{\ell\Lambda n_{\rm exc}} (r) = \int \psi^\Lambda_{n_{\rm exc}\pi}
(\bbox{r})
Y^*_{\ell\Lambda}(\hat{\bbox{r}})\,d\hat{\bbox{r}}.
\ee
Since in our calculations
the total wave function $\psi^\Lambda_{n_{\rm exc}\pi}(\bbox{r})$ is
normalized to unity, the integral
\be\label{pr}
P_{\ell}(\Lambda n_{\rm exc}) 
\equiv \int_{0}^{\infty} |R_{\ell\Lambda n_{\rm exc}} (r)|^2 r^2\,dr
\ee
represents the probability to find the partial wave $\ell$ in the state
[$n_{\rm exc} \Lambda \pi$]. Of course,
\be
\sum_{\ell}P_{\ell}(\Lambda n_{\rm exc})=1.
\ee

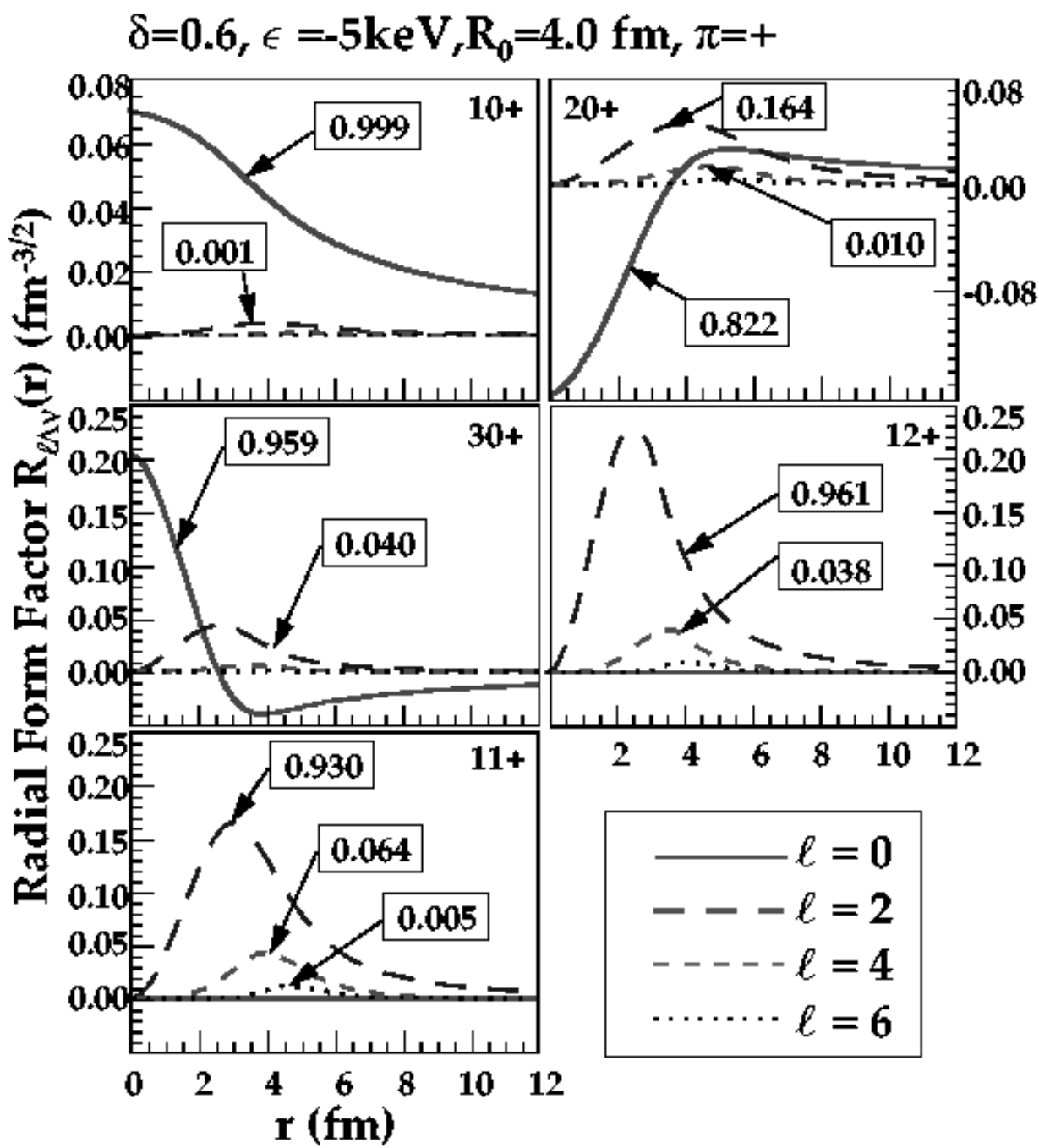
\begin{figure}[ht]
\caption {Radial form factors, Eq.~(\protect\ref{ffactors}), for several
$\pi$=+
orbitals in a superdeformed well ($\delta$=0.6, $R_0$=4\,fm)
at the 
``subthreshold'' binding
energy $\epsilon$=--5 keV.
The probability $P_{\ell}(\Lambda n_{\rm exc})$, Eq.~(\protect\ref{pr}),
is indicated.
}
\label{rff1}
\end{figure}

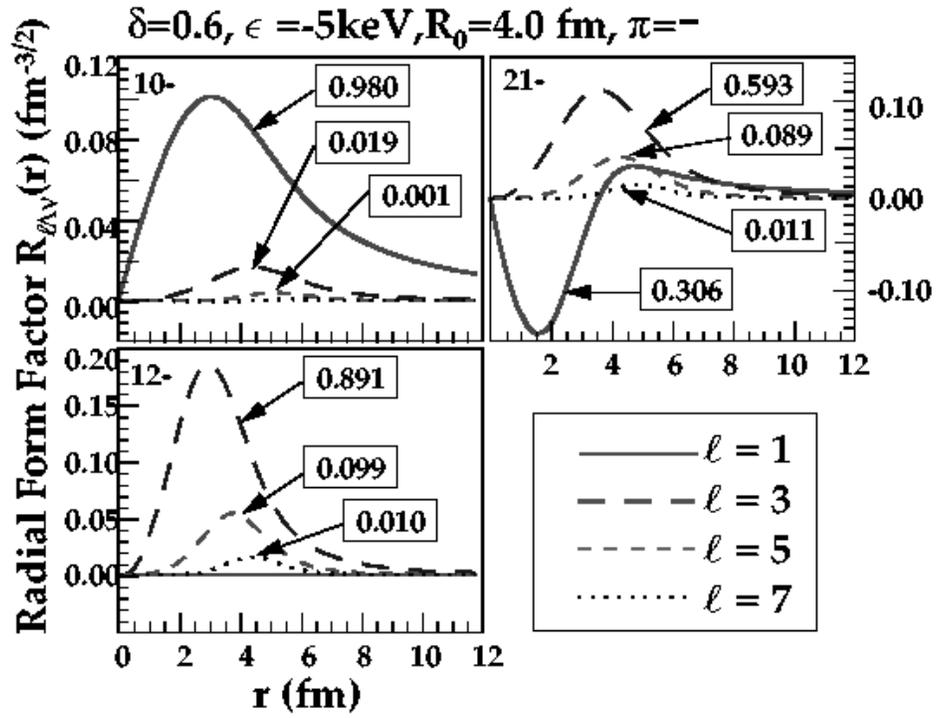
\begin{figure}[ht]
\caption{Same as in Fig.~\protect\ref{rff1} except  for
$\pi$=-- orbitals.
}
\label{rff2}
\end{figure}

Figures~\ref{rff1} and \ref{rff2} display
 the radial formfactors for several
orbitals in a superdeformed well assuming the subthreshold binding
energy of $\epsilon$=--5 keV. For the $\pi$=+, $\Lambda$=0 orbitals
(Fig.~\ref{rff1}),
the $\ell$=0 component dominates at this extremely low binding energy,
in spite of a very large deformation. 
Indeed, according to the discussion in Sect.~\ref{nint}, the 
value of $P_{\ell=0}(0 n_{\rm exc})$ approaches one at small
binding energies. In other words, the  $\pi$=+, $\Lambda$=0 halos
behave at low values of $\epsilon$ like $s$ waves.
It is interesting to note
that both the [20+] and [30+] orbitals are dominated by the $2s$ 
component; the corresponding $\ell$=0 form factors have only one node.
For the $\pi$=-- halo orbitals with
$\Lambda$=0 and 1 (Fig.~\ref{rff2}), the $p$ component
does not dominate the wave function completely (Sect.~\ref{nint}),  but
a significant excess of a $p$ wave at large distances is clearly seen.
The radial decomposition of other orbitals ($\Lambda>1$), 
 shown in Figs.~\ref{rff1} and \ref{rff2}, 
very weakly depends on binding energy; it
reflects the usual multipole mixing due to the deformed potential. 

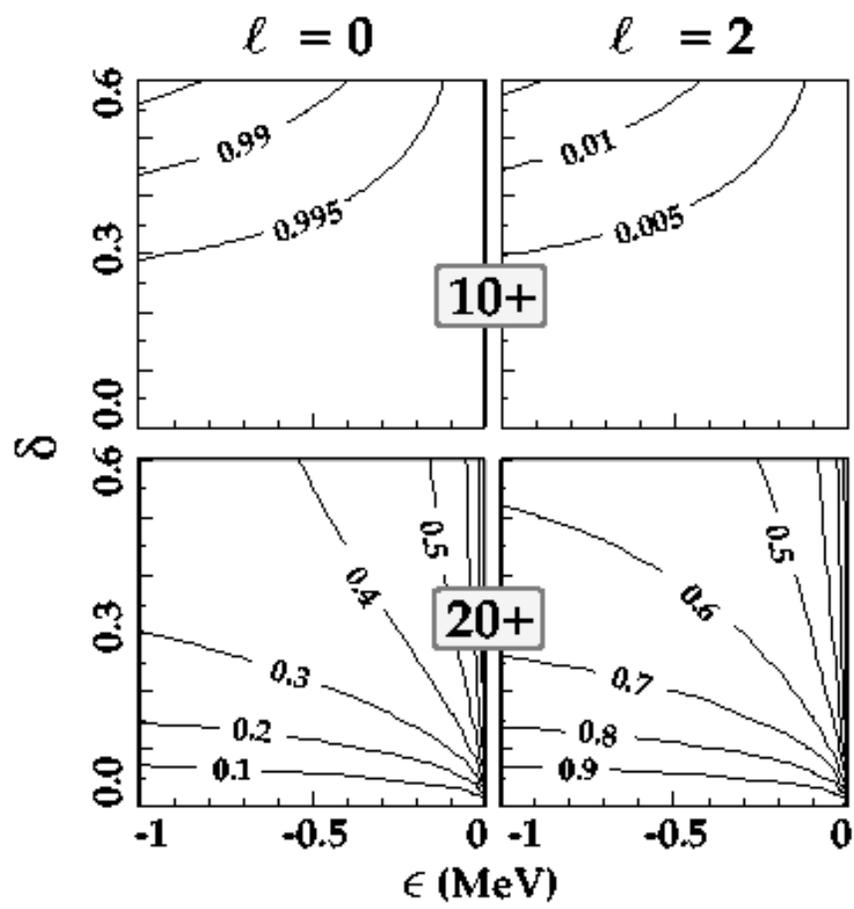
\begin{figure}[ht]
\caption {\label{ppp}
Contour maps of probabilities $P_0$ and $P_2$,
Eq.~(\protect\ref{pr}),
 for the [10+] (top) and [20+] (bottom)
levels as functions of deformation $\delta$ and binding energy
$\epsilon$. Calculations were performed for $R_0$=4\,fm.
}
\end{figure}

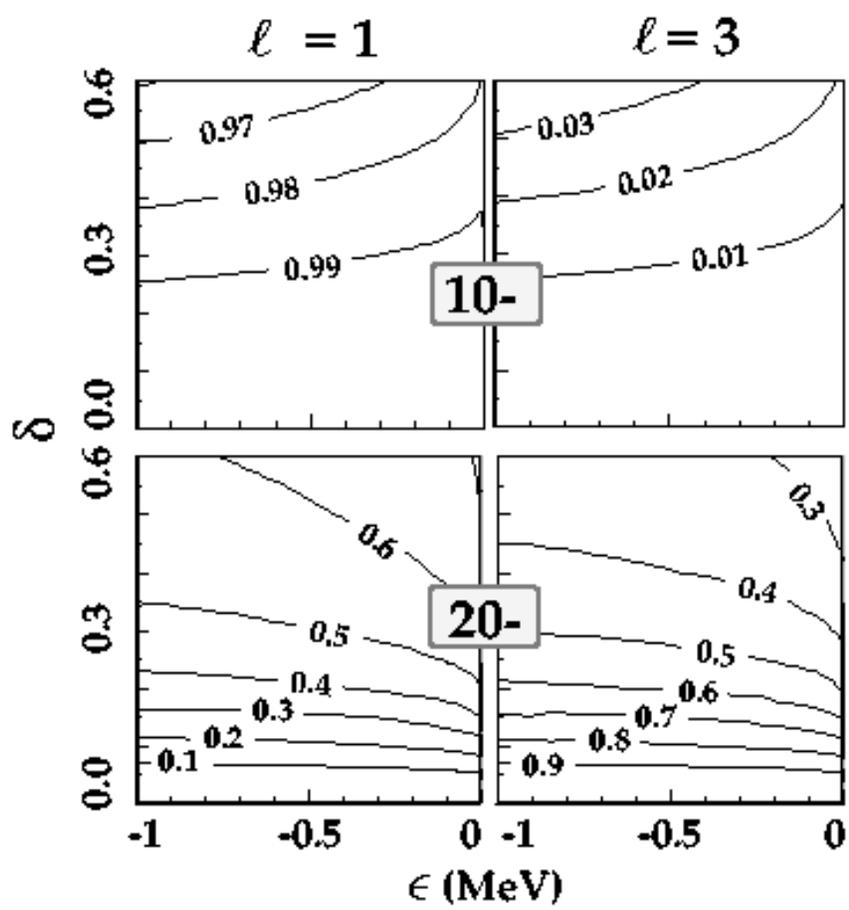
\begin{figure}[ht]
\caption {\label{ppn}
Same as in Fig.~\protect\ref{ppp} except 
for probabilities $P_1$ and $P_3$
 for the [10--] (top) and [20--] (bottom)
levels.
}
\end{figure}

The results presented  in Figs.~\ref{rff1} and   \ref{rff2}
illustrate the fact 
 that the multipole decomposition of the deformed level depends 
on {\em both}
 deformation and the  binding energy. Figures \ref{ppp}
($\pi$=+)  and \ref{ppn} ($\pi$=--)
show contour maps of $P_{\ell}(\Lambda n_{\rm exc})$ 
for the $\Lambda$=0 orbitals
as functions of $\epsilon$ and $\delta$. The structure of the [10+]
level, originating from the
spherical $1s$ state,
  is completely dominated by the $\ell$=0 component, even at very
large deformations. A rather interesting pattern is seen in the
diagram for the [20+] orbital
originating from the
spherical $1d$ state. The $\ell$=2 component dominates at low and medium
deformations, and the corresponding probability
$P_{\ell=2}$  slowly
decreases with $\delta$,  at large deformations approaching
the (constant) asymptotic limit.
 However, a similar effect, namely
the decrease of the  $\ell$=2 
component, is seen when approaching the $\epsilon$=0 threshold.
In the language of the perturbation theory \cite{[Fan61]},
 this rapid transition
comes from
the coupling to the low-energy $\ell$=0 continuum; the $\ell$=0
form factor of the [20+] orbital shows, at low values of $\epsilon$,
a one-nodal structure characteristic of the $2s$ state (see
Fig.~\ref{rff1}). At low deformations, the amplitude of the $s$ component
in the [20+] state is proportional to $\delta$.
 Remembering that the
norm, Eq.~(\ref{norm}),  behaves as $(-\epsilon)^{-1/4}$ for the
 $\ell$=0 state,
one can conclude that at low values of  $\delta$ and $\epsilon$
the contours of constant $P_{\ell=0}$ correspond to the power law
$\delta^2 \propto (-\epsilon)^{1/2}$. This result,
seen in Fig.~\ref{ppp},  tells us that
the $s$ component takes over very quickly even if  deformation $\delta$
is small. (A similar conclusion for the $1/2^+$ ground state of
$^{11}$Be has been reached in Ref.~\cite{[Rid96]}.) 

The partial-wave
probabilities calculated for the negative parity states [10--] and [20$-$]
presented in Fig.~\ref{ppn} do not 
show such a dramatic rearrangement around the threshold.
 Namely, the [10--] orbital retains its
$\ell$=1 structure in the whole deformation region considered, and
the structure of the [20--] state at large deformations 
can be viewed in terms of a mixture of $p$ and $f$ waves. In the latter case
there is a clear tendency to increase the $\ell$=1 contribution
at low binding energies, but
this  effect is much weaker compared to the [20+] case discussed above.

\subsection{Quadrupole Moments and Deformations}

Deformation properties of single-particle orbitals,
namely the intrinsic quadrupole moments 
$\langle r^2Y_{20}\rangle$  and quadrupole deformations
$\beta_2$, Eq.~(\ref{beta}), are displayed in Figs. \ref{raddp}
and \ref{raddn}
 as functions of binding energy
at $\delta$=0.6 (superdeformed shape). It is seen that the
asymptotic limits discussed in Sec.~\ref{general} are reached
at low values of $\epsilon$ in practically all cases.
In particular, the values of $\beta_2$ for the $\pi$=+, $\Lambda$=0
orbitals approach zero with $\epsilon\rightarrow 0$ [Eq.~(\ref{bpos})],
those for the $\pi$=--, $\Lambda$=0
states approach the limit of 0.63, and 
the value of $\beta_2$ for the [11$-$] orbital is close to 
the value of --0.31
[see Eq.~(\ref{bneg})].
 The only
exception is the [21--] orbital which at deformation
$\delta$=0.6 contains only
31\% of the $\ell$=1 component (see Fig.~\ref{rff2}); 
hence the positive
contribution
of the $1f$ state to $\langle r^2Y_{20}\rangle$ still dominates.

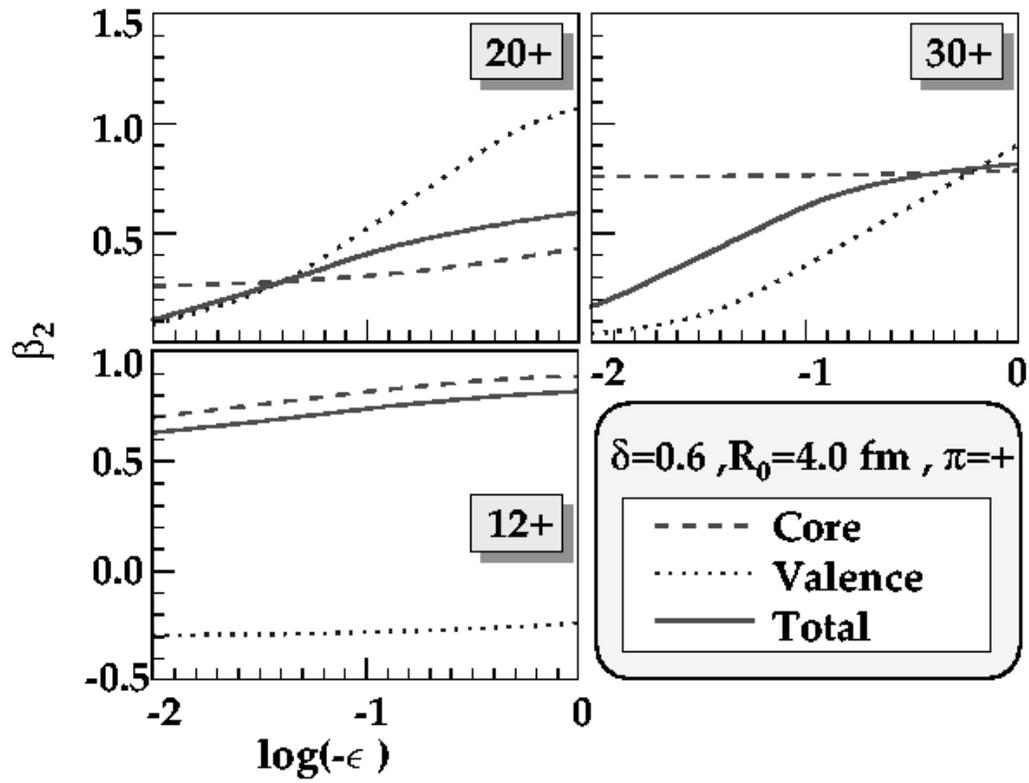
\begin{figure}[ht]
\caption {\label{b2tot1}
Quadrupole deformation $\beta_2$ of the valence orbital (dotted line), 
that of the core (dashed line), and the total quadrupole 
deformation of the system (solid line)
as a function of the binding energy of the indicated
valence orbital of positive parity.
Calculations were performed for $R_0$=4\,fm and  $\delta$=0.6.
}
\end{figure}

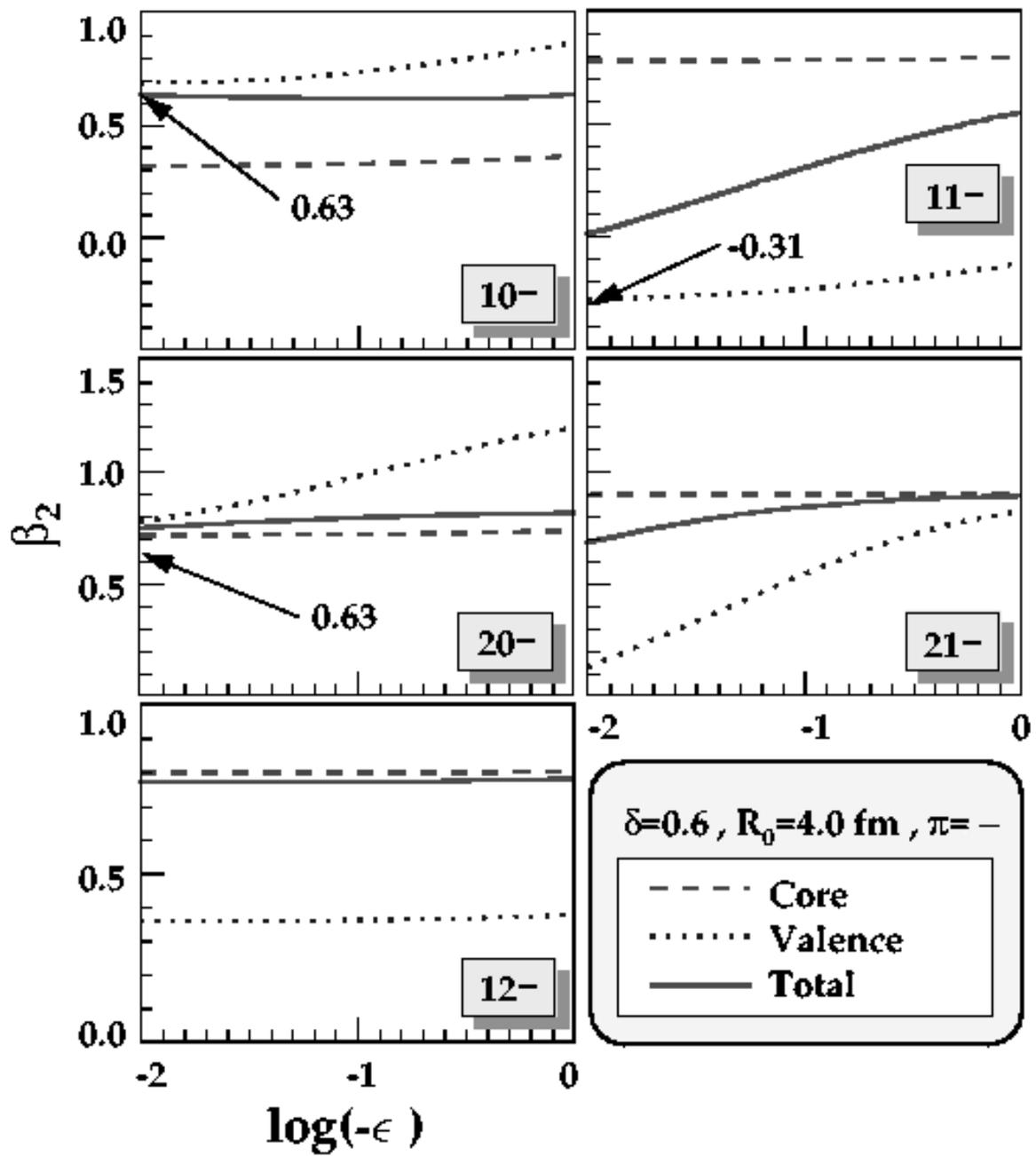
\begin{figure}[ht]
\caption {\label{b2tot2}
Same as in Fig.~\protect\ref{b2tot1} except 
for  $\pi$=-- valence orbitals.
}
\end{figure}

To illustrate
the interplay between the core deformation and that of the valence
particle
[Eq.~(\ref{b11})], we display in Figs.~\ref{b2tot1} and \ref{b2tot2}:
(i) quadrupole deformation $\beta_2$ of the valence orbital, 
(ii) quadrupole deformation of the core, and (iii) total quadrupole 
deformation of the system. Here we assume that the 
core consists of {\em all} single-particle orbitals lying energetically
below the valence orbital, and that each state (including the valence
one) is occupied by two particles.
It is convenient to rewrite Eq.~(\ref{b11}) in the form:
\be
\beta_{2,{\rm tot}} =\frac{\beta_{2,{\rm  c}}\langle r^2\rangle_{\rm c}
+\beta_{2,{\rm  v}}\langle r^2\rangle_{\rm v}}
{\langle r^2\rangle_{\rm c}+\langle r^2\rangle_{\rm v}} =
\frac{\beta_{2,{\rm  v}}+\chi\beta_{2,{\rm  c}}}{1+\chi},
\ee
where
\be
\chi \equiv \frac{\langle r^2\rangle_{\rm c}}{\langle r^2\rangle_{\rm v}}.
\ee

For the halo states, $\chi\rightarrow 0$ and
$\beta_{2,{\rm tot}}\rightarrow \beta_{2,{\rm v}}$.
The results shown in Figs.~\ref{b2tot1}
and \ref{b2tot2} nicely illustrate this behavior.
 Namely, for
very small binding energy the total deformation of the system coincides
with that of valence, {\em regardless} of the core deformation.
These results illustrate the deformation decoupling of the deformed halo
from the rest of the system. A nice example of this decoupling
has been discussed by Muta and Otsuka
\cite{[Mut95]}, who demonstrated that the halo proton in $^8$B
occupying the $1p_{3/2}$, $\Lambda$=1 weakly bound orbit  produces
the oblate density distribution which greatly reduces the large quadrupole
moment of the prolate core.

\subsection{Deformation
Softness of Halo Systems
 in the Mean-Field Calculations}\label{mfield}

The  effect of 
decoupling of the valence particles from the deformed
core in the limit of a very weak binding 
suggests that in such cases
the constrained Hartree-Fock (CHF) 
or Nilsson-Strutinsky (NS)
calculations would produce very shallow potential energy surfaces.
Indeed, in the CHF theory the nuclear Hamiltonian $H$ is minimized
under the constraint that the multipole operator that defines the intrinsic
shape has a fixed expectation value $\langle Q \rangle$=$q$. The
intrinsic wave functions are found by minimizing the Routhian
\be\label{chf}
H' = H - \beta Q,
\ee
where $\beta$ is the corresponding Lagrange multiplier.
If $Q$ is the quadrupole moment and the nucleus is weakly bound, 
then, especially
in the case of halo systems, $\langle Q \rangle$
is very sensitive to small variations in
the single-particle energy  $\epsilon_\nu$  of the last occupied
single-particle orbital.
In particular, for the
$\pi$=-- halos, $\langle Q \rangle$
can take practically any value without changing the HF energy
$\langle H \rangle$. This means that the numerical procedure
used for  searching for the self-consistent solution can be 
rather susceptible
to  uncontrolled   variations of $q$ with $\epsilon_\nu$.

In the Nilsson-Strutinsky calculations, the bulk part of the binding
energy comes from the 
 the macroscopic energy
$E_{\rm macro}$. Commonly used is the Yukawa-plus-exponential
macroscopic energy formula \cite{[Kra79]} which
accounts for the surface thickness. The corresponding
generalized surface energy reads
\be\label{krappe}
E_s = -\frac{c_s}{8\pi^2r_0^2a^3}\int\int_V\left(\frac{\sigma}{a}-2\right)
\frac{e^{-\sigma/a}}{\sigma}d^3\bbox{r}\,d^3\bbox{r}',
\ee
where $c_s$ is the surface-energy coefficient, 
$R_0$=$r_0 A^{1/3}$, $a$ is the surface diffuseness parameter, 
$\sigma=|\bbox{r}-\bbox{r}'|$,
and $V$ denotes the volume enclosed by the deformed
nuclear surface. The latter has been
 defined in our study by means of the 
axial multipole expansion in terms of
deformation
parameters $\beta_\lambda$:
 \be\label{radius1}
R(\Omega ) = c(\beta )R_0\left[
1 + \sum_{\lambda} \beta_{\lambda}Y_{\lambda 0}
(\Omega )\right]
 \ee
with $c(\beta )$ being determined from the volume-conservation condition.

As demonstrated in Ref. \cite{[Kra79]}, 
for small deformations
the generalized surface energy, Eq.~(\ref{krappe}),
is given to second order by
\be\label{krappe1}
E_s = E_s({\rm sph}) + \sum_{\lambda} c_\lambda(\zeta) \beta^2_\lambda,
\ee
where the expansion coefficients $c_\lambda$ solely depend
on the dimensionless parameter 
\be\label{xi}
\zeta=\frac{R_0}{a}=\frac{r_0 A^{1/3}}{a}. 
\ee
It can be shown \cite{[Jon96]} that the function $c_\lambda(\zeta)$
becomes negative below the critical value of $\zeta_c$ which is roughly
proportional to the multipolarity $\lambda$. Consequently,
the generalized surface energy is stable to $\beta_\lambda$ if
$\zeta > \zeta_c \approx 0.8 \lambda$, or
\be\label{krappe2}
a\lambda < R_0/0.8.
\ee
According to Eq.~(\ref{krappe2}), 
for a given nucleus
{\em both} large multipolarity 
{\em and} large
diffuseness can trigger the shape instability.
This conclusion also holds for the finite-range droplet model mass
formula \cite{[Jon96]}.

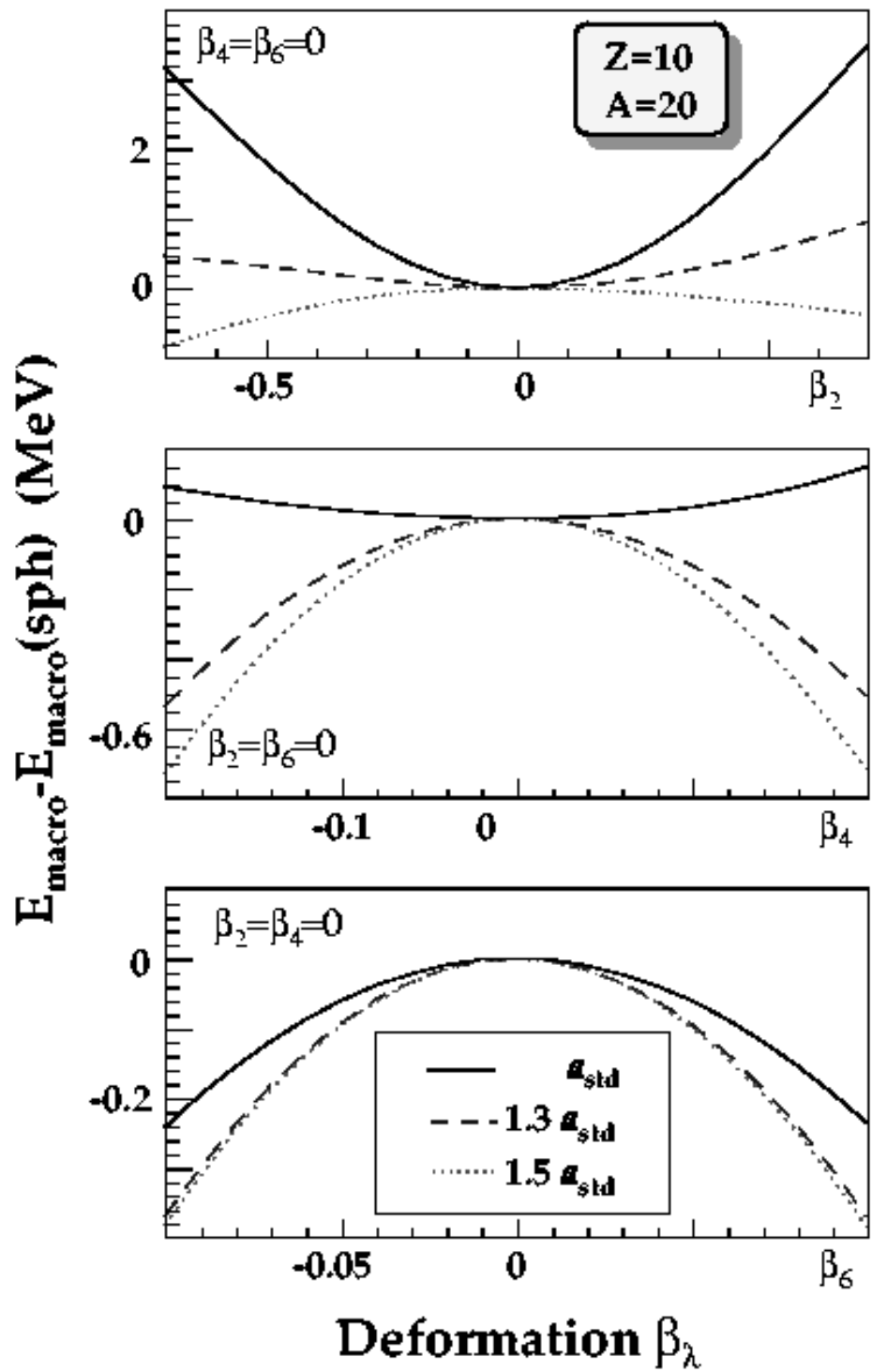
\begin{figure}[ht]
\caption {The macroscopic energy,
$E_{\rm macro}$, for $Z$=$N$=10
(normalized to zero at
spherical shape) given by
the  Yukawa-plus-exponential  mass formula
as a function of $\beta_2$ (top), $\beta_4$ (middle), and 
$\beta_6$ (bottom) for three values of surface diffuseness:
$a_{\rm std}$=0.68\,fm (solid line),
$a$=1.3$a_{\rm std}$ (dashed line),
and $a$=1.5$a_{\rm std}$ (dotted line). 
It is seen that the macroscopic energy develops
deformation instability 
with {\em  both} increasing 
diffuseness and the
multipolarity of deformation.
}
\label{LD}
\end{figure}

The weakly bound neutron-rich nuclei and halo systems are characterized
by very diffused density distributions. For instance, it has been
predicted in Ref. \cite{[Dob94]} that the average diffuseness
in neutron drip-line nuclei can increase by as much as 50\% as compared
to the standard value representative of nuclei around the beta stability
line. The effect of the  large diffuseness on 
the macroscopic energy  is illustrated
in Fig. \ref{LD}, which displays $E_{\rm macro}$
with the parameters of Ref. \cite{[Mol88]}
 for a light 
$A$=20 nucleus as a function of deformations $\beta_2$, $\beta_4$, and 
$\beta_6$. 
Since at low deformations
different multipolarities  are decoupled, Eq.~(\ref{krappe1}),
they can be varied separately and the remaining ones
are put to zero. The calculations are performed for three
values of $a$.
It is seen that the general rule given by Eq.~(\ref{krappe2})
holds. Namely, for larger values of $a$ and $\lambda$
the macroscopic energy 
becomes unstable
to shape deformation, mainly due to instability of $E_s$
(for very light nuclei the effect of the Coulomb term is much weaker).
Interestingly, the effect is fairly pronounced even for quadrupole
distortions; the potential energy curve becomes unstable to $\beta_2$
already for $a$=1.3$a_{\rm std}$. The above results indicate
that in the microscopic-macroscopic approach both single-particle
and macroscopic energy become extremely shallow to deformation
for weakly bound systems.

\section{Conclusions}\label{conclusions}

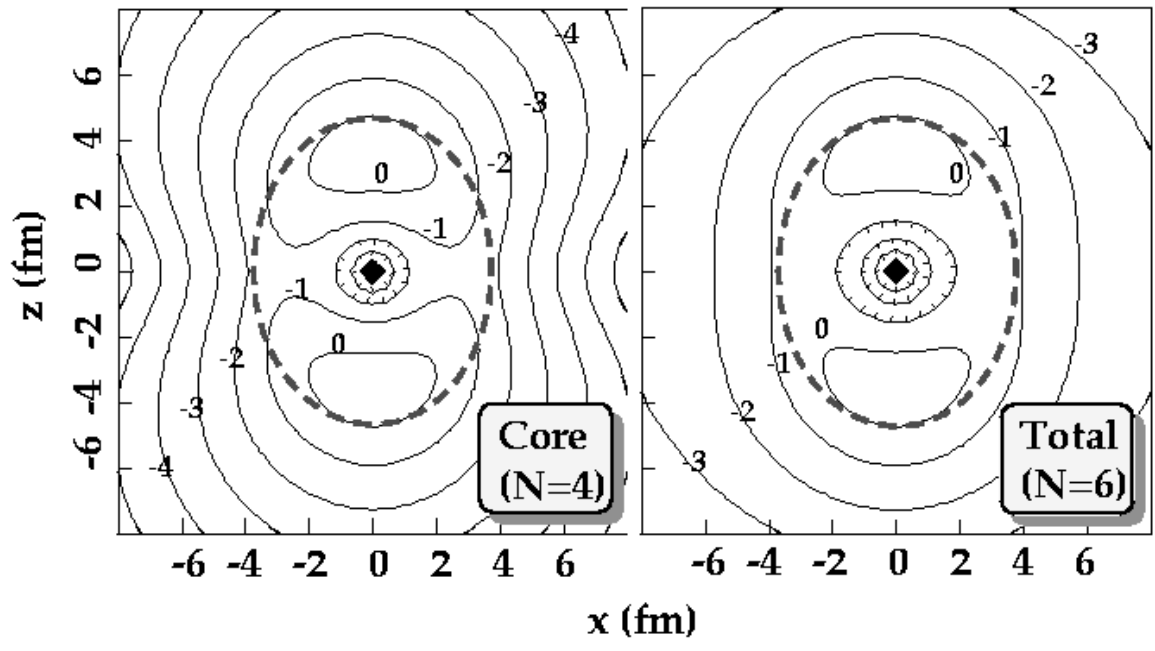
\begin{figure}[ht]
\caption {Contour plot (in logarithmic scale) of
the  $\rho r^2$  distribution
in the  ($x$, $z$)-plane. Here, $\rho$ denotes
the single-particle density, i.e., 
$\rho(\protect\bbox{r})$=$\sum_{\nu-{\rm occ.}}
|\psi_\nu(\protect\bbox{r})|^2$.
The deformation of the spheroidal well is
assumed to be  $\delta$=0.2
and the corresponding prolate shape is
indicated  by a dashed line. Left:
 distribution for a core consisting of $N$=4 particles
(counting from the bottom of the well). Right:
 $\rho r^2$ distribution for the  
$N$=6 system. The two
valence orbitals occupy the weakly bound $\epsilon$=--5\,keV
[11--] orbital.
}
\label{density}
\end{figure}

In the
limit of  very weak binding,
the geometric  interpretation of shape deformation is lost.
  Consider, e.g., a deformed
core with a prolate  deformation and a weakly-bound halo neutron
in a negative-parity orbital.
According to the discussion above, the total quadrupole moment
of the system diverges at the limit of 
vanishing binding (i.e., $\langle r^2Y_{20}\rangle$
 can take {\em any}
value).   
On the other hand, 
depending on the geometry of the valence orbital,
the total quadrupole deformation of the (core+valence) system
is consistent with 
a superdeformed shape  ($\pi$=--, $\Lambda$=0 halo) or
oblate shape ($\pi$=--, $\Lambda$=1 halo).
For a $\pi$=+ halo, the quadrupole moment is finite but
$\beta_2$ approaches zero.
In the language of the self-consistent mean-field theory, this result
reflects the extreme softness of the system to the quadrupole distortion.
Figure~\ref{density} shows an example of such a situation:
The two valence particles occupying the weakly bound
[11--] orbital give rise to an oblate deformation of the system,
in spite of the prolate deformation of the core
and the prolate shape of the underlying spheroidal well 
($\delta$=0.2).

Shape
deformation is an extremely  powerful concept provided that the 
nuclear surface can be properly defined. However,
 for very diffused and spatially extended systems
the geometric interpretation of multipole moments and deformations
is lost.

The presence of the spatially extended  neutron halo  
gives rise to  the 
presence  of low-energy isovector modes in neutron-rich nuclei.
The deformation decoupling of the halo
implies that the nuclei close to the neutron drip line are
excellent candidates for isovector quadrupole deformations,
 with different 
quadrupole deformations for protons and neutrons. Such  nuclei
 are expected to have a very interesting 
rotational  behavior and unusual magnetic properties. 
For instance, the rotational features  of such systems (moments of inertia,
$B(E2)$ values, $g$-factors) should be solely determined
by the deformed core.

An example of the above scenario has been predicted in the self-consistent
calculations 
for the neutron-rich sulfur isotopes
performed
 using Skyrme
Hartree-Fock and  relativistic mean field 
methods \cite{[Wer94a],[Wer96]}.
When approaching
the neutron drip line, the calculated values of $\beta_2$ 
for neutrons are systematically
smaller than those
 of the proton distribution.
This example illustrates once again that
in the drip-line nuclei, due to spatially extended wave functions,
the ``radial" contribution to the quadrupole moment might be
as important as the ``angular" part. 

Finally, it is interesting to note that the anisotropic (non-spherical)
halo systems have been investigated in molecular physics. A direct
molecular analogy of a 
quadrupole-deformed halo nucleus is the electron weakly
bound by the quadrupole moment of the neutral symmetric molecule such
as CS$_2$ \cite{[Com96]}.

\section*{Acknowledgments}

This work 
has been supported by the U.S. Department of
Energy through Contract No. DE-FG05-93ER40770.
Oak Ridge National
Laboratory is managed for the U.S. Department of Energy by Lockheed
Martin Energy Research Corp. under Contract No.
DE-AC05-96OR22464.
The Joint Institute for Heavy Ion
 Research has as member institutions the University of Tennessee,
Vanderbilt University, and the Oak Ridge National Laboratory; it
is supported by the members and by the Department of Energy
through Contract No. DE-FG05-87ER40361 with the University
of Tennessee.  

\appendix
\section{The computation of single-particle wave functions}
\label{appA}

The amplitudes $A_{n_{\rm exc}\Lambda l}$ and $B_{n_{\rm exc}\Lambda l}$ in
 Eq.~(\ref{waveint}) can be found following the method outlined in
Ref.~\cite{[Mer94]}.
The matching conditions for 
the internal and external wave functions (\ref{waveint})
at $\xi=\xi_0$
lead to the following set of 
equations:
\be\label{bound1}
\sum_{l}^{\infty}A_{n_{\rm exc}\Lambda l}R^{(1)}_{\Lambda l}
(c_{\rm int},\xi_0)
S_{\Lambda l}(c_{\rm int},\eta) = 
\sum_{l}^{\infty}B_{n_{\rm exc}\Lambda l}
R^{(3)}_{\Lambda l}(ic_{\rm ext},\xi_0)
S_{\Lambda l}(ic_{\rm ext},\eta)
\ee
and
\be\label{bound2}
\sum_{l}^{\infty}A_{n_{\rm exc}\Lambda l}\left(\frac{dR^{(1)}_{\Lambda l}
(c_{\rm int},\xi)}
{d\xi}\right)_{\xi_0}S_{\Lambda l}(c_{\rm int},\eta)
=  \sum_{ l}^{\infty}B_{n_{\rm exc}\Lambda l}
\left(\frac{dR^{(3)}_{\Lambda l}
(ic_{\rm ext},\xi)}{d\xi}\right)_{\xi_0}S_{\Lambda l}
(ic_{\rm ext},\eta).
\ee
The matching conditions [Eqs.~(\ref{bound1}) and (\ref{bound2})]
should hold for any value of $\eta$. To eliminate  this
degree of freedom, one can take advantage of the fact that
  the angular
spheroidal functions $S_{\Lambda l}(c,\eta)$ form a complete orthogonal 
set in the interval $-1\leq \eta\leq 1$. By multiplying
both sides of Eqs.~(\ref{bound1}) and (\ref{bound2}) by
$S^*_{\Lambda l'}(ic_{\rm ext},\eta)$
and integrating  over  $\eta$,
one obtains the matrix equation for $A_{n_{\rm exc}\Lambda l}$:
\be\label{boundf}
\sum_{l}^{\infty}A_{n_{\rm exc}\Lambda l}M_{ll'}
 = 0~~\text{for all}~l',
\ee
where
\begin{eqnarray}\label{boundmat}
M_{ll'}  \equiv  
\left\{\left(\frac{dR^{(1)}_{\Lambda l}(c_{\rm int},\xi)}
{d\xi}\right)_{\xi_0}R^{(3)}_{\Lambda l'}
(ic_{\rm ext},\xi_0) 
\right.& - & \left.\left(\frac{dR^{(3)}_{\Lambda l'}
(ic_{\rm ext},\xi)}{d\xi}\right)_{\xi_0}R^{(1)}_{\Lambda l}
(c_{\rm int},\xi_0)\right\}\nonumber\\
 & \times & \langle S_{\Lambda l'}(ic_{\rm ext})|
S_{\Lambda l}(c_{\rm int})\rangle.
\end{eqnarray}

The eigenenergies $\epsilon_{n_{\rm exc}\Lambda\pi}$ 
and the corresponding amplitudes  $A_{n_{\rm exc}\Lambda l}$
are found from the condition that
$\det(M)$=0.

\section{Separation of the Schr\"odinger Equation}
\label{appB}

In the limit of very small binding energy and/or
large deformations,
 particular attention should be paid to the 
numerical method
employed for determining the separation constant 
$\lambda_{\Lambda l}$ [Eq.~(\ref{schrodsep})]
and amplitudes $d^{\Lambda l}_k$ appearing in 
Eqs.~(\ref{angwave1}) and (\ref{radwave}).

  The separation constant  $\lambda_{\Lambda l}$ and 
coefficients $d^{\Lambda l}_k(c)$ can be obtained from the
 three-term recurrence
formula \cite{[Mor53]}:
\begin{eqnarray}\label{threerec}
\left[\frac{k(k-1)}{(2k+2\Lambda-1)(2k+2\Lambda-3)}
\right]d^{\Lambda l}_{k-2} +
\left[\frac{(k+2\Lambda+1)(k+2\Lambda+2)}
{(2k+2\Lambda+3)(2k+2\Lambda+5)}\right]
d^{\Lambda l}_{k+2}\nonumber\\
+\left[\frac{2(k+\Lambda)(k+\Lambda+1)-
2\Lambda^2-1}{(2k+2\Lambda+3)(2k+2\Lambda-1)}+
\frac{(k+\Lambda)(k+\Lambda+1)-\lambda_{\Lambda l}}
{c^2}\right]d^{\Lambda l}_k
=0.
\end{eqnarray}	
It is interesting to note that for $c\rightarrow 0$ [the last
term in  Eq.~(\ref{threerec}) dominates] $l$=$\ell$,
 the separation constant
becomes $\lambda_{\Lambda l}=\ell(\ell+1)$, and the only 
nonvanishing coefficient corresponds to $k$=$\ell$--$\Lambda$. 
That is, the wave functions
(\ref{angwave1}) and (\ref{radwave}) are those of the 
{\em spherical} well. Indeed, according to
Eq.~(\ref{c22}), the limit of  $c\rightarrow 0$ is reached
at the spherical shape ($a\rightarrow 0$). However, the limit
of $c_{\rm ext}\rightarrow 0$ is also approached at {\em deformed
shapes} if the binding energy is small. 

By defining  
\be
\alpha_k\equiv d_{k-\Lambda}/
\sqrt{\frac{2k+1}{2}\frac{(k-\Lambda)!}{(k+\Lambda)!}},
\ee 
the matrix equation
(\ref{threerec})
can be reduced to the Hermitian eigenvalue problem \cite{[Mer94]}
\begin{eqnarray}\label{eig2}
\lambda_{\Lambda l}(c)\alpha_k & = &
\frac{c^2}{2k+3}\sqrt{\frac{(k+\Lambda+1)(k+\Lambda+2)
(k-\Lambda+1)(k-\Lambda+2)}{(2k+1)(2k+5)}}\alpha_{k+2}\nonumber\\&
&
+\left\{k(k+1)+c^2\frac{2k(k+1)-2\Lambda^2-1}
{(2k+3)(2k-1)}\right\}\alpha_k\nonumber\\&
&
+\frac{c^2}{2k-1}\sqrt{\frac{(k+\Lambda-1)
(k+\Lambda)(k-\Lambda-1)(k-\Lambda)}{(2k-3)(2k+1)}}\alpha_{k-2}.
\end{eqnarray}

Another way  of computing $\lambda_{\Lambda l}$
and  $d^{\Lambda l}_k$ is to employ the recurrence relations for
the ratio $d_k/d_{k-2}$ \cite{[Abr65],[Lit56]}:
\be\label{ratio}
\frac{d_{k+2}}{d_k} + \beta_k + 
\gamma_k\left(\frac{d_k}{d_{k-2}}\right)^{-1}=0,
\ee
where $\beta_k$ and $\gamma_k$
can be expressed through  factors appearing
in Eq.~(\ref{threerec}).

The technique based on the recurrence relation (\ref{ratio})
is useful when calculating radial wave functions at large deformations.
 In general, the
coefficients $d_k$ are largest in magnitude 
at  $k= l-\Lambda$, and
quickly decrease with $k$.
On the other hand, the magnitudes 
 of spherical Bessel functions of the third kind,
$h_k(ic\xi)$, 
rapidly increase with  $k$. Conseqently,
 the product
$d^{\Lambda l}_k(c)h_{k+\Lambda}(ic\xi)$ becomes numerically unstable at
large values of $k$ if $d_k$ and $h_{k+\Lambda}$ are computed separately.
This problem can be overcome
by writing  the recurrence relation  (\ref{ratio}) in the form
\be\label{ratio1}
\left(\frac{d_{k+2}h_{k+\Lambda+2}}{d_kh_{k+\Lambda}}\right)
\left(\frac{h_{k+\Lambda+2}}{h_{k+\Lambda}}\right)^{-1}
+ \beta_k +
\gamma_k\left(\frac{d_kh_{k+\Lambda}}{d_{k-2}h_{k+\Lambda-2}}
\right)^{-1}\left(\frac{h_{k+\Lambda}}{h_{k+\Lambda-2}}\right)
= 0
\ee
and employing the property of  Bessel functions $h_k(iz)$
\be\label{bessr2}
\frac{h_{k+2}}{h_k}+1+\frac{2k+3}{z}\left[
\frac{2k+1}{z}+\frac{z}{2k-1}\left\{1+\left(\frac{h_k}
{h_{k-2}}\right)^{-1}\right\}\right]=0.
\ee


\end{document}